\documentclass[11pt]{article}
\usepackage{graphicx}
\usepackage{latexsym,graphicx}
\usepackage{float}
\usepackage{amsmath}
\usepackage{amsfonts}
\usepackage{amssymb}
\usepackage{epstopdf}
\usepackage{color}
\usepackage{cite}
\DeclareGraphicsExtensions{.eps}
\usepackage[left=2.5cm,top=2.5cm,right=2.5cm,bottom=2.5cm]{geometry}

\catcode `\@=11
\catcode `\@=12
\begin{document}
	\begin{center}
		\large{\bf{Cosmological implications and causality in $f(R, L_{m}, T)$ gravity theory with observational constraints}} \\
		\vspace{5mm}
		\normalsize{ Dinesh Chandra Maurya$^{1, a}$, Javlon Rayimbaev$^{2, b,c,d,e}$, Inomjon Ibragimov$^{3, f}$, Sokhibjan Muminov$^{4, g}$}\\
		\vspace{5mm}
		\normalsize{$^{a}$Centre for Cosmology, Astrophysics and Space Science, GLA University, Mathura-281 406,
		Uttar Pradesh, India.}\\
		\vspace{2mm}
		\normalsize{$^{b}$Institute of Fundamental and Applied Research, National Research University TIIAME, Kori Niyoziy 39, Tashkent 100000, Uzbekistan.}\\
		\vspace{2mm}
		\normalsize{$^{c}$University of Tashkent for Applied Sciences, Str. Gavhar 1, Tashkent 100149, Uzbekistan.}\\
		\vspace{2mm}
		\normalsize{$^{d}$Urgench State University, Kh. Alimjan Str. 14, Urgench 221100, Uzbekistan.}\\
		\vspace{2mm}
		\normalsize{$^{e}$Shahrisabz State Pedagogical Institute, Shahrisabz Str. 10, Shahrisabz 181301, Uzbekistan.}\\
		\vspace{2mm}
		\normalsize{$^{f}$Kimyo International University in Tashkent, Shota Rustaveli street 156, Tashkent 100121, Uzbekistan.}\\
		\vspace{2mm}
		\normalsize{$^{g}$Mamun university, Bolkhovuz Street 2, Khiva 220900, Uzbekistan.}\\
		\vspace{2mm}
		$^{1}$E-mail:dcmaurya563@gmail.com \\
		\vspace{2mm}
        $^{2}$E-mail:javlon@astrin.uz \\
        \vspace{2mm}
        $^{3}$E-mail:i.ibragimov@kiut.uz \\
        \vspace{2mm}
        $^{4}$E-mail:sokhibjan.muminov@mamunedu.uz \\
	\end{center}
	\vspace{5mm}
	\begin{abstract}		
     In the generalized matter-geometry coupling theory, we investigate the physical characteristics and causality of some new cosmological models for a flat, homogeneous, and isotropic spacetime filled with stiff, radiation, dust, and curvature fluid sources. We obtain a particular cosmological model corresponding to each source fluid, called Models I, II, III, and IV, respectively. We make observational constraints on each model using the joint analysis of $31$ Cosmic Chronometer (CC) Hubble dataset and $1701$ Pantheon+SH0ES datasets to estimate the current values of model parameters. Using these statistical results, we have analyzed the information criteria, effective EoS parameter, causality of the models, and viability of this generalized gravity theory. Subsequently, we investigate the effective equation of state and deceleration parameter for each model. We found that all models in the late-time universe exhibit transit-phase acceleration, and Models I and II show both the early as well as late-time accelerating phase of the expanding universe. We found the current values of the deceleration parameter in the range $-0.8857\le q_{0}\le-0.4279$ with transition redshift $0.4867\le z_{t}\le0.839$ and the effective EoS parameter in the range $-0.9238\le\omega_{eff}\le-0.6186$. We analyzed the square sound speed condition $c_{s}^{2}\le c^{2}$ for each model.
	\end{abstract}
	\smallskip
	\vspace{5mm}
	\vspace{1cm}
	
	\section{Introduction}
	
    The discovery of the accelerating phase of the expanding universe \cite{ref1,ref2,ref3,ref4,ref5}, the cosmological constant problem \cite{ref6,ref7,ref8,ref9,ref10,ref11}, and the fundamental laws of physics force to modify Einstein's general theory of gravity (GR). Although, several modifications to Einstein's general relativity theory of gravity (GR) have been introduced in the literature, but we will only discuss some of these modifications within the Riemannian framework here, since, we are interested in investigating the generalized matter-geometry coupling gravity theory. In 1970, Buchdahl \cite{ref12} suggested a wider version of GR theory by changing the Lagrangian $L=R$ to $L=f(R)$ in the Einstein-Hilbert action. He focused on an open universe model that changed between non-singular states, and this idea was later studied in \cite{ref13,ref14,ref15,ref16}. Recently, cosmologists found that the universe is expanding faster than they thought. They use this theory to explain this and to solve problems with the cosmological constant \cite{ref17, ref18, ref19, ref20, ref21, ref22, ref23, ref24, ref25, ref26, ref27, ref28, ref29, ref30, ref31}. The modified $f(R)$-gravity theory presents initial cosmological models that include both cosmic phenomena, inflation, and cosmic speed-up in the expansion of the universe. Recently, we have investigated several cosmological models in the context of a Riemannian framework to explain the accelerating scenarios and evolution of expanding universe properties in \cite{ref32,ref33,ref34,ref35,ref36,ref37,ref38,ref39,ref40}.\\
    
    The extended $f(R)$-gravity theories are introduced in \cite{ref41,ref42,ref43} which create a relationship between matter and curvature by including them externally using the expression $S=\int[f_{1}(R)+(1+\lambda f_{2}(R))L_{m}]\sqrt{-g}d^{4}x$. The extra relationship between matter and curvature results in additional forces. This theory is later improved by Harko \cite{ref44}, and this time, he proposed the matter-geometry relationship with action $S=\int[\frac{1}{2}f_{1}(R)+G(L_{m}) f_{2}(R)]\sqrt{-g}d^{4}x$. After that, this idea was developed into the $f(R, L_{m})$ gravity theory by Harko and Lobo \cite{ref45}. In this improved form of matter-geometry coupling theory, the Lagrangian function $f$ is an arbitrary function of the Ricci curvature scalar $R$ and the matter Lagrangian $L_{m}$. This matter-geometry coupling theory is used by several cosmologists \cite{ref46,ref47,ref48,ref49,ref50,ref51,ref52,ref53,ref54,ref55,ref56,ref57} to explore the late-time cosmic accelerating scenarios and observable universe physical phenomena in recent literature. We have, also recently investigated transit cosmological models in the $f(R,L_{m})$ gravity theory, using observational constraints in \cite{ref58,ref59,ref60,ref61,ref62}.\\
    
    Subsequently, a further study \cite{ref63} approached matter-geometry coupling in a different form called $f(R,T)$ gravity theory. At this time, the $f$ is an arbitrary function of the Ricci scalar $R$ and trace $T$ of the energy-momentum tensor $T_{ij}$. Furthermore, this geometry creates a non-minimal correlation between the trace of the matter-energy-momentum tensor. In \cite{ref64}, a thermodynamic interpretation of matter-geometry coupling models is given, and \cite{ref65} investigated cosmic evolution of $f(R,T)$ gravity models with collisional matter. \cite{ref66} has presented a scalar tensor approach of $f(R,T)$ gravity models. Stability of $f(R,T)$ gravity models is investigated in \cite{ref67}, and \cite{ref68} has studied the in-viability of $f(R,T)$ gravity theory. In \cite{ref69,ref70,ref71}, some more cosmological implications and late-time cosmic acceleration in $f(R,T)$ gravity models are investigated. Recently, we have investigated the late-time scenarios of the cosmological models in $f(R,T)$ gravity \cite{ref72,ref73,ref74,ref75,ref76}. Particle creation \cite{ref77} and variable Chaplygin gas \cite{ref78} models are investigated in $f(R,T)$ gravity theory.\\
    
    Further unification of $f(R)$, $f(R,L_{m})$, and $f(R,T)$ gravity theories is proposed by Haghani and Harko \cite{ref79} in the generalized form of matter-geometry coupling, which is referred to as $f(R,L_{m},T)$ gravity. The main objective of developing this theory is to integrate modified gravity theories in Riemannian geometry, including the $f(R)$, $f(R,L_{m})$, and $f(R,T)$ gravity theories. Although theories such as $f(R)$-gravity, $f(T)$-gravity, and $f(Q)$-gravity have their own advantages in describing the universe's history, our specific focus is on the more comprehensive and less investigated theory of $f(R,L_{m},T)$ gravity. This theory guarantees that the modification is substantial for all types of matter fields. Physical factors are the reason for including this background in the examination of cosmological theories. Recently, \cite{ref80} has discussed the energy conditions in $f(R,L_{m},T)$ gravity theory, and \cite{ref81} investigated dusty universe in $f(R,L_{m},T)$ gravity to achieve accelerating cosmological scenarios using observational constraints. The causality and its violation in $f(R,L_{m},T)$ gravity theory is discussed in \cite{ref82}. Recently, we have developed some constrained accelerating cosmological models and its stability analysis in \cite{ref83,ref84}. Here, we investigate physical characteristics and causality of the expanding universe in this unified $f(R,L_{m},T)$ gravity in the presence of perfect fluid with constant equation of state (EoS) parameter. We use the flat, homogeneous, and isotropic Friedmann-Lema\^{\i}tre-Robertson-Walker (FLRW) spacetime universe because, in this case, we are not to fix the properties openness or closedness of the universe \cite{ref85}.\\
    
    We organize the present research paper into six sections. A brief introduction of the development of generalized matter-geometry coupling theory is given in Sect.\,1. The field equations in this generalized gravity theory are derived in Sect.\,2. We obtain the cosmological solutions in Sect.\,3, while the observational constraints made on the models are in Sect.\,4. The result discussions with causality analysis are in Sect.\,5, and in the last Sect.\,6, concluding remarks are given.

\section{Field equations in $f(R,L_{m},T)$ gravity theory}

We consider the action for generalized matter-geometry coupling $f(R,L_{m},T)$ gravity theory in the following form \cite{ref79}
\begin{equation}\label{eq1}
I=\frac{1}{16\pi}\int{f(R,L_{m},T)\sqrt{-g}d^{4}x}+\int{L_{m}\sqrt{-g}d^{4}x},
\end{equation}
where $f(R,L_{m},T)$ is an arbitrary function of the Ricci scalar, $R$, the trace $T$ of the stress-energy momentum tensor of the matter, $T_{ij}$, and of the matter Lagrangian density $L_{m}$. We define the stress-energy momentum tensor of the matter as \cite{ref86}
\begin{equation}\label{eq2}
  T_{ij}=-\frac{2}{\sqrt{-g}}\frac{\delta(\sqrt{-g}L_{m})}{\delta{g^{ij}}},
\end{equation}
and its trace by $T=g^{ij}T_{ij}$, respectively. By assuming that the Lagrangian density $L_{m}$ of matter depends only on
the metric tensor components $g_{ij}$, and not on its derivatives, we obtain
\begin{equation}\label{eq3}
  T_{ij}=g_{ij}L_{m}-2\frac{\partial{L_{m}}}{\partial{g^{ij}}}.
\end{equation}
By varying the action $I$ of the gravitational field with respect to the metric tensor components $g^{ij}$ provides the following relationship
\begin{equation}\label{eq4}
  \delta{I}=\int\left[{f_{R}\delta{R}+f_{T}\frac{\delta{T}}{\delta{g^{ij}}}\delta{g^{ij}}+f_{L_{m}}\frac{\delta{L_{m}}}{\delta{g^{ij}}}\delta{g^{ij}}-\frac{1}{2}g_{ij}f\delta{g^{ij}}}-8\pi T_{ij}\delta{g^{ij}}\right]\sqrt{-g}d^{4}x,
\end{equation}
where we have denoted $f_{R}=\partial{f}/\partial{R}$, $f_{T}=\partial{f}/\partial{T}$ and $f_{L_{m}}=\partial{f}/\partial{L_{m}}$, respectively. For the variation of the Ricci scalar, we obtain
\begin{equation}\label{eq5}
  \delta{R}=\delta(g^{ij}R_{ij})=R_{ij}\delta{g^{ij}}+g^{ij}(\nabla_{\lambda}\delta\Gamma^{\lambda}_{ij}-\nabla_{j}\delta\Gamma^{\lambda}_{i\lambda}),
\end{equation}
where $\nabla_{\lambda}$ is the covariant derivative with respect to the symmetric connection $\Gamma$ associated to the metric $g$. The variation of the Christoffel symbols yields
\begin{equation}\label{eq6}
  \delta{\Gamma^{\lambda}_{ij}}=\frac{1}{2}g^{\lambda\alpha}(\nabla_{i}\delta{g_{j\alpha}}+\nabla_{j}\delta{g_{\alpha i}}-\nabla_{\alpha}\delta{g_{ij}}),
\end{equation}
and the variation of the Ricci scalar provides the expression
\begin{equation}\label{eq7}
  \delta{R}=R_{ij}\delta{g^{ij}}+g_{ij}\Box\delta{g^{ij}}-\nabla_{i}\nabla_{j}\delta{g^{ij}}.
\end{equation}
Therefore, for the variation of the action of the gravitational field, we obtain
\begin{equation}\label{eq8}
 \delta{R}=\int\left[f_{R}R_{ij}+(g_{ij}\Box-\nabla_{i}\nabla_{j})f_{R}+f_{T}\frac{\delta(g^{\alpha\beta}T_{\alpha\beta})}{\delta{g^{ij}}}+f_{L_{m}}\frac{\delta{L_{m}}}{\delta{g^{ij}}}-\frac{1}{2}g_{ij}f-8\pi T_{ij}\right]\delta{g^{ij}}\sqrt{-g}d^{4}x.
\end{equation}
We define the variation of $T$ with respect to the metric tensor as
\begin{equation}\label{eq9}
  \frac{\delta(g^{\alpha\beta}T_{\alpha\beta})}{\delta{g^{ij}}}=T_{ij}+\Theta_{ij},
\end{equation}
where
\begin{equation}\label{eq10}
  \Theta_{ij}=g^{\alpha\beta}\frac{\delta T_{\alpha\beta}}{\delta{g^{ij}}}=L_{m}g_{ij}-2T_{ij},
\end{equation}
for the perfect-fluid matter source.\\
Taking $\delta{I}=0$, we obtain the field equations of $f(R,L_{m}, T)$ gravity model as
\begin{equation}\label{eq11}
  f_{R}R_{ij}-\frac{1}{2}[f-(f_{L_{m}}+2f_{T})L_{m}]g_{ij}+(g_{ij}\Box-\nabla_{i}\nabla_{j})f_{R}=\left[ 8\pi+\frac{1}{2}(f_{L_{m}}+2f_{T})\right]T_{ij}.
\end{equation}
By contracting Eq.~(\ref{eq11}) gives the following relation between the Ricci scalar $R$ and the trace $T$ of the stress-energy tensor,
\begin{equation}\label{eq12}
f_{R}R-2[f-(f_{L_{m}}+2f_{T})L_{m}]+3\Box f_{R}=\left[ 8\pi+\frac{1}{2}(f_{L_{m}}+2f_{T})\right]T,
\end{equation}
where we have denoted $\Theta=\Theta_{i}^{i}$.\\
The problem of the perfect fluids, described by an energy density $\rho$, pressure $p$ and four-velocity $u^{i}$ is more complicated, since there is no unique definition of the matter Lagrangian. However, in the present study we assume that the stress-energy tensor of the matter is given by
\begin{equation}\label{eq13}
  T_{ij}=(\rho+p)u_{i}u_{j}+pg_{ij},
\end{equation}
for the flat FLRW homogeneous and isotropic spacetime metric 
\begin{equation}\label{eq14}
	ds^{2}=-dt^{2}+a(t)^{2}(dx^{2}+dy^{2}+dz^{2}),
\end{equation}
and the matter Lagrangian can be taken as $L_{m}=-\rho$. The four-velocity $u_{i}$ satisfies the conditions $u_{i}u^{i}=-1$ and $u^{i}\nabla_{j}u_{i}=0$, respectively.\\
The energy-conservation equation is obtained as
\begin{equation}\label{eq15}
	\dot{\rho}+3H(\rho+p)=\frac{(\rho+L_{m})(\dot{f}_{T}+\frac{1}{2}\dot{f}_{L_{m}})-\frac{1}{2}f_{T}(\dot{T}-2\dot{L}_{m})}
	{8\pi+f_{T}+\frac{1}{2}f_{L_{m}}}.
\end{equation}
	or
	\begin{equation}\label{eq16}
		\dot{\rho}+3H(\rho+p)=D_{\mu}u^{\mu}
	\end{equation}
	The source term $D_{\mu}u^{\mu}$ represents the energy production or destruction within the system. The conservation of total energy in a gravitational system is attained alone when the condition $D_{\mu}u^{\mu}=0$ is fulfilled at every point in spacetime. If $D_{\mu}u^{\mu} \neq 0$, an energy transfer process or particle formation occurs within the system.\\
	Comparing Eqs.\,\eqref{eq15} and \eqref{eq16}, we obtain
	\begin{equation}\label{eq17}
		D_{\mu}u^{\mu}=\frac{(\rho+L_{m})(\dot{f}_{T}+\frac{1}{2}\dot{f}_{L_{m}})-\frac{1}{2}f_{T}(\dot{T}-2\dot{L}_{m})}
		{8\pi+f_{T}+\frac{1}{2}f_{L_{m}}}.
	\end{equation}
	Equations \eqref{eq16} and \eqref{eq17} indicate that the source term $D_{\mu}u^{\mu}\ne0$, signifying the formation or destruction of matter within the system. In the present study, we assume $D_{\mu}u^{\mu}=0$ and solve the field equations for different fluid sources which satisfies this condition.

\section{Cosmological solutions}

To investigate the cosmological properties of the above-proposed modified gravity, we consider the following form of the matter-geometry coupling function $f(R,L_{m},T)$ with matter Lagrangian $L_{m}=-\rho$, (as suggested in \cite{ref79}),
\begin{equation}\label{eq18}
  f(R,L_{m},T)= R+\mu\,T\,L_{m}-\nu,
\end{equation}
where $\mu$ and $\nu$ are the arbitrary constants.\\
Then
\begin{equation}\label{eq19}
	f_{R}=1,~~~~ f_{T}=\mu\,L_{m}, ~~~~ f_{L_{m}}=\mu\,T.
\end{equation}
By applying \eqref{eq18} in \eqref{eq11}, we get the following field equations:

\begin{equation}\label{eq20}
	R_{ij}-\frac{1}{2}R\,g_{ij}=(8\pi+\frac{\mu}{2}\,T+\mu\,L_{m})T_{ij}-(\mu\,L_{m}^{2}+\frac{\nu}{2})g_{ij}.
\end{equation}
Now, using \eqref{eq13} and \eqref{eq14} in \eqref{eq20}, we obtain the following equations:

\begin{equation}\label{eq21}
	3H^{2}=8\pi\rho-\frac{\mu}{2}\rho^{2}+\frac{3\mu}{2}\rho p+\frac{\nu}{2},
\end{equation}
\begin{equation}\label{eq22}
	2\dot{H}+3H^{2}=-8\pi p+\frac{3\mu}{2}(\rho-p)p+\mu\rho^{2}+\frac{\nu}{2}.
\end{equation}
From Eqs.~\eqref{eq15} and \eqref{eq18}, we obtain the equation of continuity as follows
\begin{equation}\label{eq23}
		\dot{\rho}+3H(\rho+p)=\frac{\mu\rho(\dot{\rho}+3\dot{p})}{16\pi-3\mu(\rho-p)}.
\end{equation}
Since the $f(R,L_{m},T)$ theory is a non-conservative theory and hence, generally, it does not satisfy the energy conservation equation i.e., $D_{\mu}u^{\mu}\ne0$. Therefore, we consider the conserved matter fluid sources that satisfies the continuity equations $\dot{\rho}+3H(1+\omega_{i})\rho=0$ for $i=s, r, m, c$, which stand for stiffness, radiation, dust, and curvature fluids, respectively. Now, we solve the highly nonlinear Eqs.~\eqref{eq21} and \eqref{eq22} for different phase of perfect fluid ($p=\omega_{i}\rho$) as stiff, radiation, dust and curvature in four subsections as follows:

\subsection{Model-I}

To investigate the physical characteristics and accelerating scenarios in the context of barotropic fluid $\omega\ge0$, we consider the perfect fluid source as stiff fluid ($\omega=1$) for which the above Eqs.~\eqref{eq21} and \eqref{eq22} reduced to
\begin{equation}\label{eq24}
	3H^{2}=8\pi\rho+\mu\rho^{2}+\frac{\nu}{2},
\end{equation}
\begin{equation}\label{eq25}
	2\dot{H}+3H^{2}=-8\pi\rho+\mu\rho^{2}+\frac{\nu}{2}.
\end{equation}
The energy density for stiff fluid is obtained after solving the continuity equation $\dot{\rho}+3H(1+\omega_{s})\rho=0$, as
\begin{equation}\label{eq26}
	\rho=\rho_{s0}(1+z)^{6},
\end{equation}
where $\rho_{s0}$ is the current value of energy density and $1+z=\frac{a_{0}}{a}$ as given in \cite{ref87}.\\
Now, we derive the Hubble function from Eqs.~\eqref{eq24} and \eqref{eq26} as
\begin{equation}\label{eq27}
	H(z)=H_{0}\sqrt{\Omega_{s}(1+z)^{6}+\Omega_{\mu}(1+z)^{12}+\Omega_{\nu}},
\end{equation}
where $\Omega_{s}=\frac{8\pi\rho_{s0}}{3H_{0}^{2}}$, $\Omega_{\mu}=\frac{\mu\rho_{s0}^{2}}{3H_{0}^{2}}$ and $\Omega_{\nu}=\frac{\nu}{6H_{0}^{2}}$. At $z=0$, $\Omega_{s}+\Omega_{\mu}+\Omega_{\nu}=1$.\\
The physical behaviour of model is characterized by the equation of state parameter $\omega$, and hence, we derive the effective equation of state (EoS) parameter $\omega_{eff}$ is from Eqs.\,\eqref{eq24} and \eqref{eq25} as
\begin{equation}\label{eq28}
	\omega_{eff}=-1+\frac{2\Omega_{s}(1+z)^{6}}{\Omega_{s}(1+z)^{6}+\Omega_{\mu}(1+z)^{12}+\Omega_{\nu}}.
\end{equation}
The phase of the expanding universe is characterized by the deceleration parameter $q(z)$, and the deceleration parameter $q(z)$ is obtained as
\begin{equation}\label{eq29}
	q(z)=\frac{1}{2}+\frac{3}{2}\frac{\Omega_{s}(1+z)^{6}-\Omega_{\mu}(1+z)^{12}-\Omega_{\nu}}{\Omega_{s}(1+z)^{6}+\Omega_{\mu}(1+z)^{12}+\Omega_{\nu}}.
\end{equation}

\subsection{Model-II}

This model is derived for the perfect fluid as radiation fluid ($\omega=\frac{1}{3}$) and $p=\frac{1}{3}\rho$, to get the characteristics of the model under radiation fluid sources, and hence, the above Eqs.~\eqref{eq21} and \eqref{eq22} reduced to
\begin{equation}\label{eq30}
	3H^{2}=8\pi\rho+\frac{\nu}{2},
\end{equation}
\begin{equation}\label{eq31}
	2\dot{H}+3H^{2}=-\frac{8\pi}{3}\rho+\frac{4\mu}{3}\rho^{2}+\frac{\nu}{2}.
\end{equation}
The energy density for radiation fluid source is obtained after solving the continuity equation $\dot{\rho}+3H(1+\omega_{r})\rho=0$ as given by
\begin{equation}\label{eq32}
	\rho=\rho_{r0}(1+z)^{4},
\end{equation}
where $\rho_{r0}$ is the current value of the energy density.\\
We derive the Hubble function from the Eqs.~\eqref{eq30} and \eqref{eq32} as
\begin{equation}\label{eq33}
	H(z)=H_{0}\sqrt{\Omega_{r}(1+z)^{4}+\Omega_{\nu}},
\end{equation}
where $\Omega_{r}=\frac{8\pi\rho_{r0}}{3H_{0}^{2}}$ and $\Omega_{\nu}=\frac{\nu}{6H_{0}^{2}}$. At $z=0$, $\Omega_{r}+\Omega_{\nu}=1$. The effective EoS parameter $\omega_{eff}$ for this model, is derived as
\begin{equation}\label{eq34}
	\omega_{eff}=-1+\frac{4}{3}\frac{\Omega_{r}(1+z)^{4}-\Omega_{\mu}(1+z)^{8}}{\Omega_{r}(1+z)^{4}+\Omega_{\nu}},
\end{equation}
where $\Omega_{\mu}=\frac{\mu\rho_{r0}^{2}}{3H_{0}^{2}}$. The deceleration parameter is obtained as
\begin{equation}\label{eq35}
	q(z)=\frac{1}{2}+\frac{1}{2}\frac{\Omega_{r}(1+z)^{4}-4\Omega_{\mu}(1+z)^{8}-\Omega_{\nu}}{\Omega_{r}(1+z)^{4}+\Omega_{\nu}}.
\end{equation}

\subsection{Model-III}

In this model, we consider the perfect fluid source as dust ($\omega=0$) and $p=0$ to get the characteristics of the cosmological models in this modified generalized theory of gravity, and hence, the above Eqs.~\eqref{eq21} and \eqref{eq22} reduced to
\begin{equation}\label{eq36}
	3H^{2}=8\pi\rho-\frac{\mu}{2}\rho^{2}+\frac{\nu}{2},
\end{equation}
\begin{equation}\label{eq37}
	2\dot{H}+3H^{2}=\mu\rho^{2}+\frac{\nu}{2}.
\end{equation}
The energy density for dusty fluid source is obtained from the equation of continuity $\dot{\rho}+3H(1+\omega_{m})\rho=0$ as given by
\begin{equation}\label{eq38}
	\rho=\rho_{m0}(1+z)^{3},
\end{equation}
where $\rho_{m0}$ is the current value of energy density.\\
The Hubble function is derived from Eqs.~\eqref{eq36} and \eqref{eq38} as
\begin{equation}\label{eq39}
	H(z)=H_{0}\sqrt{\Omega_{m}(1+z)^{3}+\Omega_{\mu}(1+z)^{6}+\Omega_{\nu}},
\end{equation}
where $\Omega_{m}=\frac{8\pi\rho_{m0}}{3H_{0}^{2}}$, $\Omega_{\mu}=-\frac{\mu\rho_{m0}^{2}}{6H_{0}^{2}}$ and $\Omega_{\nu}=\frac{\nu}{6H_{0}^{2}}$. At $z=0$, $\Omega_{m}+\Omega_{\mu}+\Omega_{\nu}=1$. The effective EoS parameter $\omega_{eff}$ for this model is derived as
\begin{equation}\label{eq40}
	\omega_{eff}=-1+\frac{\Omega_{m}(1+z)^{3}+3\Omega_{\mu}(1+z)^{6}}{\Omega_{m}(1+z)^{3}+\Omega_{\mu}(1+z)^{6}+\Omega_{\nu}}.
\end{equation}
The deceleration parameter is obtained as
\begin{equation}\label{eq41}
	q(z)=\frac{1}{2}+\frac{3}{2}\frac{2\Omega_{\mu}(1+z)^{6}-\Omega_{\nu}}{\Omega_{m}(1+z)^{3}+\Omega_{\mu}(1+z)^{6}+\Omega_{\nu}}.
\end{equation}

\subsection{Model-IV}

The present model is derived for perfect fluid source as curvature ($\omega=-\frac{1}{3}$) and $p=-\frac{1}{3}\rho$, to investigate the cosmological properties of the universe, and hence, the Eqs.~\eqref{eq21} and \eqref{eq22} reduced to
\begin{equation}\label{eq42}
	3H^{2}=8\pi\rho-\mu\rho^{2}+\frac{\nu}{2},
\end{equation}
\begin{equation}\label{eq43}
	2\dot{H}+3H^{2}=\frac{8\pi}{3}\rho+\frac{\mu}{3}\rho^{2}+\frac{\nu}{2}.
\end{equation}
The equation of continuity is obtained as
\begin{equation}\label{eq44}
	\dot{\rho}+2H\rho=0.
\end{equation}
Hence, the energy density for curvature is obtained as
\begin{equation}\label{eq45}
	\rho=\rho_{c0}(1+z)^{2},
\end{equation}
where $\rho_{c0}$ is the current value of energy density\\
We derive the Hubble function from Eqs.~\eqref{eq42} and \eqref{eq45} as
\begin{equation}\label{eq46}
	H(z)=H_{0}\sqrt{\Omega_{c}(1+z)^{2}-\Omega_{\mu}(1+z)^{4}+\Omega_{\nu}},
\end{equation}
where $\Omega_{c}=\frac{8\pi\rho_{c0}}{3H_{0}^{2}}$, $\Omega_{\mu}=\frac{\mu\rho_{c0}^{2}}{3H_{0}^{2}}$ and $\Omega_{\nu}=\frac{\nu}{6H_{0}^{2}}$. At $z=0$, $\Omega_{c}-\Omega_{\mu}+\Omega_{\nu}=1$. The effective EoS parameter $\omega_{eff}$ for this model is derived as
\begin{equation}\label{eq47}
	\omega_{eff}=-1+\frac{2}{3}\frac{\Omega_{c}(1+z)^{2}-4\Omega_{\mu}(1+z)^{4}}{\Omega_{c}(1+z)^{2}-\Omega_{\mu}(1+z)^{4}+\Omega_{\nu}}.
\end{equation}
The deceleration parameter is obtained as
\begin{equation}\label{eq48}
	q(z)=\frac{1}{2}-\frac{1}{2}\frac{\Omega_{c}(1+z)^{2}+\Omega_{\mu}(1+z)^{4}+3\Omega_{\nu}}{\Omega_{c}(1+z)^{2}-\Omega_{\mu}(1+z)^{4}+\Omega_{\nu}}.
\end{equation}

\section{Observational Constraints}

This section aims to constrain the free model parameters of our derived models \eqref{eq27}, \eqref{eq33}, \eqref{eq39}, and \eqref{eq46} utilizing accessible datasets concerning the evolution of the expanding universe. We utilize $31$ non-correlated cosmic chronometer (CC) Hubble data points \cite{ref88,ref89}, obtained by the differential age technique, along with $1701$ data points from the Pantheon$+$SH0ES sample of SNe Ia \cite{ref90,ref91,ref92,ref93,ref94}. We analyze the observational datasets employing the conventional Bayesian methodology and utilize the Monte Carlo Markov Chain (MCMC) technique to ascertain the posterior variations of the parameters. We also employ the emcee software for MCMC analysis, which is available at no cost \cite{ref95}. Moreover, the probability function $\mathcal{L}\propto{e^{-\chi^{2}/2}}$ yields optimal values for the free parameters by maximizing $\mathcal{L}$ or reducing the $\chi^{2}$-value.\\

\subsection{Hubble datasets}

We commence with the $31$ Hubble data points of the Hubble parameter $H(z)$ associated with different redshifts $z$ from the cosmic chronometer (CC) sample. The CC dataset of $H(z)$ is derived using the differential age method. The Hubble parameter may directly signify the rate of cosmic expansion in the universe. The decision to employ CC data is mostly contingent upon age difference measurements between two passively evolving galaxies that arose simultaneously, characterized by a small redshift interval. This interval facilitates the calculation of $\delta{z}/\delta{t}$. CC data has demonstrated greater reliability than previous methods reliant on absolute age determinations for galaxies \cite{ref92}. To assess the model's goodness of fit with the data, we utilize the following $\chi^{2}$ formula defined as:
\begin{equation}\label{eq49}
	\chi_{CC}^{2}=\sum_{i=1}^{i=N}\frac{[H_{ob}(z_{i})-H_{th}(\phi, z_{i})]^{2}}{\sigma_{H(z_{i})}^{2}}.
\end{equation}
The symbol $\phi$ represents the collection of model parameters for a particular model. $H_{ob}$ refers to the observed values, while $H_{th}$ represents the theoretical values of the Hubble function. $\sigma_{H(z_{i})}$ denotes the standard deviations associated with the Hubble data points. Lastly, $N$ indicates the total number of data points used in the analysis.\\

\subsection{Pantheon$+$SH0ES data}
We utilize the Pantheon+SH0ES data \cite{ref90,ref91} in the range of redshift $0.00122\le z \le2.26137$ extracted from the 18 different surveys. This latest dataset of Pantheon+SH0ES contains $1701$ light curves of $1550$ distinct types of SNe Ia, and this sample is an extended version of the Pantheon sample at the low redshifts \cite{ref96}. In the Pantheon sample, we find a degeneracy between the Hubble constant $H_{0}$ and apparent magnitude $M$ ; hence, we were unable to constrain both parameters $H_{0}$ and $M$ together. But the Pantheon+SH0ES dataset breaks this degeneracy between $H_{0}$ and $M$, and this allows for the estimation of $H_{0}$ and $M$ together \cite{ref92}. The Pantheon+ sample also incorporates SH0ES \cite{ref93} Cepheid host distances to constrain the absolute magnitude parameter $M$. In this analysis, we use the $\chi^{2}$ function as
\begin{equation}\label{eq50}
	\chi^{2}_{pp}=\Delta\mu^{T}C^{-1}\Delta\mu
\end{equation}
where $C^{-1}$ shows the inverse covariance matrix generated by incorporating systematic and statistical covariance matrices \cite{ref94}, and the $i$th element of vector $\Delta\mu$ is determined by \cite{ref90,ref92}
\begin{equation}\label{eq51}
	\Delta\mu_{i}=\left\{ \begin{array}{ll}
		m_{i}-M-\mu_{i}^{Cepheid} &, \text{if } i\in \text{Cepheid hosts} \\
		m_{i}-M-\mu_{model}(z_{i}) &, \text{otherwise }
	\end{array} \right.
\end{equation}
where $\mu_{i}^{Cepheid}$ is the distance modulus of the Cepheid host for the $i$th SNe Ia \cite{ref94}. The standardized and adjusted values of $m_b^{corr}$ magnitudes are denoted by $m_{i}$ which are available in the `m\_b\_corr' column of the Pantheon+ data \cite{ref94}. The theoretical apparent magnitude $m(z_{i})$ is defined as
\begin{equation}\label{eq52}
	m(z_{i})=M+ 5~\log_{10}\left(\frac{d_{l}(z_{i})}{Mpc}\right)+25.
\end{equation}
The theoretical distance modulus $\mu_{model}(z_{i})$ is defined as
\begin{equation}\label{eq53}
	\mu_{model}(z_{i})=5\log_{10}{d_l(z_i)+25},
\end{equation}
where $d_l$ is the luminosity distance defined as
\begin{equation}\label{eq54}
	d_l=c(1+z_{hel})\int_{0}^{z_{HD}}\frac{dz'}{H(z')}
\end{equation}
Here, $z_{HD}$ and $z_{hel}$ are extracted from the `zHD' and `zhel' column of the dataset \cite{ref94}. $z_{HD}$ denotes the redshift in the CMB frame with peculiar velocity corrections, and $z_{hel}$ is the heliocentric redshift of the supernovae.\\
We also constrain the parameter $M$ in our analysis, given its potential influence in a cosmological model. It is interesting to note that with the use of SH0ES Cepheid distance values available with the `IS\_CALIBRATOR' column in data \cite{ref94}, the degeneracy between $H_{0}$ and $M$ may be eliminated in this analysis. This is an intriguing and useful aspect of the Pantheon+SHOES dataset as compared to the Pantheon data.\\
To get constrained values of model parameters involved in the Hubble function of derived Models I, II, III, and IV, using joint analysis of CC and CC$+$Pantheon+SH0ES datasets, we use the following $\chi^{2}$ formula:
\begin{equation}\label{eq55}
	\chi^{2}=\chi^{2}_{CC}+\chi^{2}_{pp}
\end{equation}
\begin{table}[H]
	\centering
	\begin{tabular}{|c|c|c|c|}
		\hline
  Model     &Parameter            & Prior              & Value\\
		\hline
		    &$H_{0}$              & $(50, 100)$        & $73.54\pm0.96$\\
    I       & $\Omega_{s}$        & $(0, 0.05)$        & $0.0381\pm0.0025$\\
		    &$\Omega_{\mu}$       & Fixed              & $0.0000001$\\
		    & $M$                 & $(-20, -18)$       & $-19.317\pm0.027$\\
		    &$\chi_{min.}^{2}$    & $-$                & $1771.3013$\\
		\hline
		    &$H_{0}$              & $(40, 100)$        & $72.24\pm0.92$\\
	II      & $\Omega_{r}$        & $(0, 0.3)$         & $0.1699\pm0.0090$\\
		    & $M$                 & $(-20, -18)$       & $-19.307\pm0.026$\\
		    &$\chi_{min.}^{2}$    & $-$                & $1581.2280$\\
		\hline
		    &$H_{0}$              & $(10, 100)$        & $72.01\pm0.92$\\
   III      & $\Omega_{m}$        & $(0, 0.9)$         & $0.315\pm0.018$\\
		    &$\Omega_{\mu}$       & $(0, 0.005)$       & $0.00120_{-0.0012}^{+0.00027}$\\
		    & $M$                 & $(-20, -18)$       & $-19.295\pm0.026$\\
		    &$\chi_{min.}^{2}$    & $-$                & $1547.7011$\\
		\hline
		    &$H_{0}$              & $(10, 100)$        & $71.87\pm0.92$\\
    IV      & $\Omega_{c}$        & $(0.01, 0.99)$     & $0.490\pm0.077$\\
		    &$\Omega_{\mu}$       & $(-0.2, 0.01)$     & $-0.041\pm0.020$\\
		    & $M$                 & $(-20, -18)$       & $-19.289\pm0.026$\\
		    &$\chi_{min.}^{2}$    & $-$                & $1544.6908$\\
		\hline
		    &$H_{0}$              & $(40, 90)$         & $72.02\pm0.91$\\
$\Lambda$CDM& $\Omega_{m}$        & $(0.01, 0.6)$      & $0.316\pm0.020$\\
		    & $M$                 & $(-20, -18)$       & $-19.295\pm0.025$\\
		    &$\chi_{min.}^{2}$    & $-$                & $1547.7061$\\
		\hline
	\end{tabular}
	\caption{The MCMC estimates with priors in different models.}\label{T1}
\end{table}
\begin{figure}[H]
	\centering
	a.\includegraphics[width=8cm,height=8cm,angle=0]{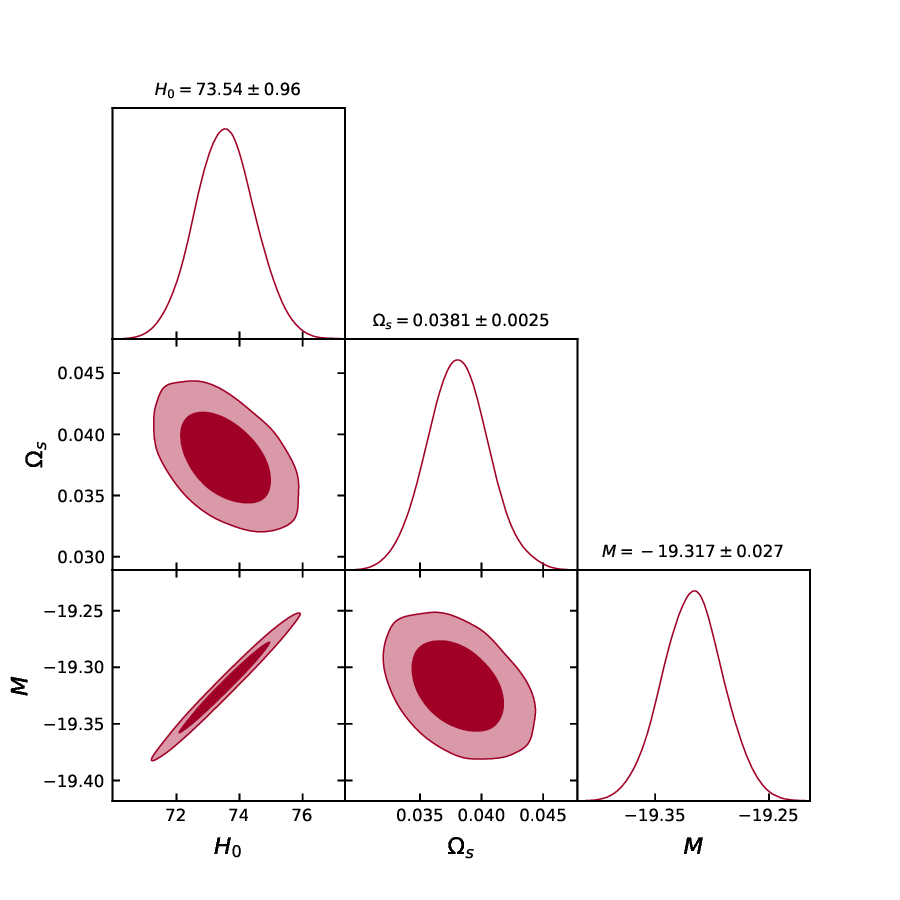}
	b.\includegraphics[width=8cm,height=8cm,angle=0]{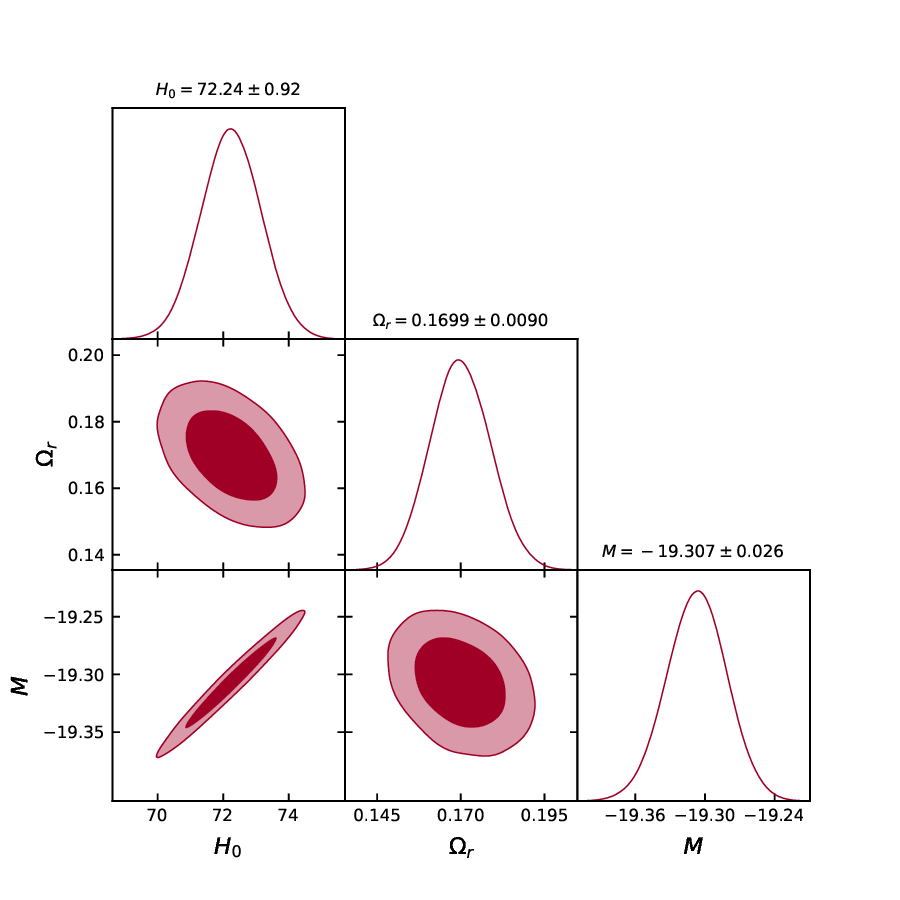}
	\caption{The contour plots of $H_{0}$, $\Omega_{s}$, $M$ and the contour plots of $H_{0}$, $\Omega_{r}$, $M$, respectively, for joint analysis of CC$+$Pantheon+SH0ES.}
\end{figure}
\begin{figure}[H]
	\centering
	a.\includegraphics[width=8cm,height=8cm,angle=0]{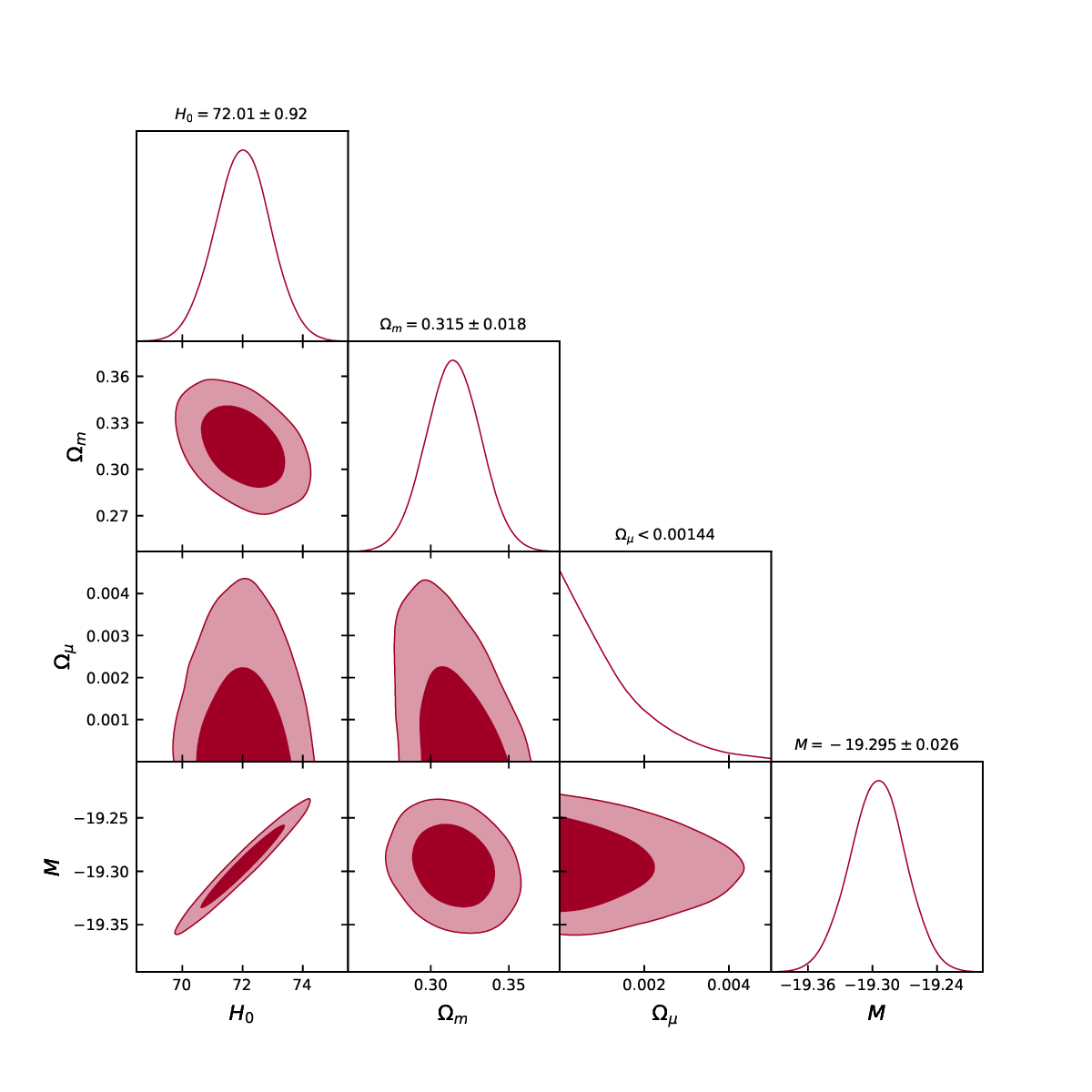}
	b.\includegraphics[width=8cm,height=8cm,angle=0]{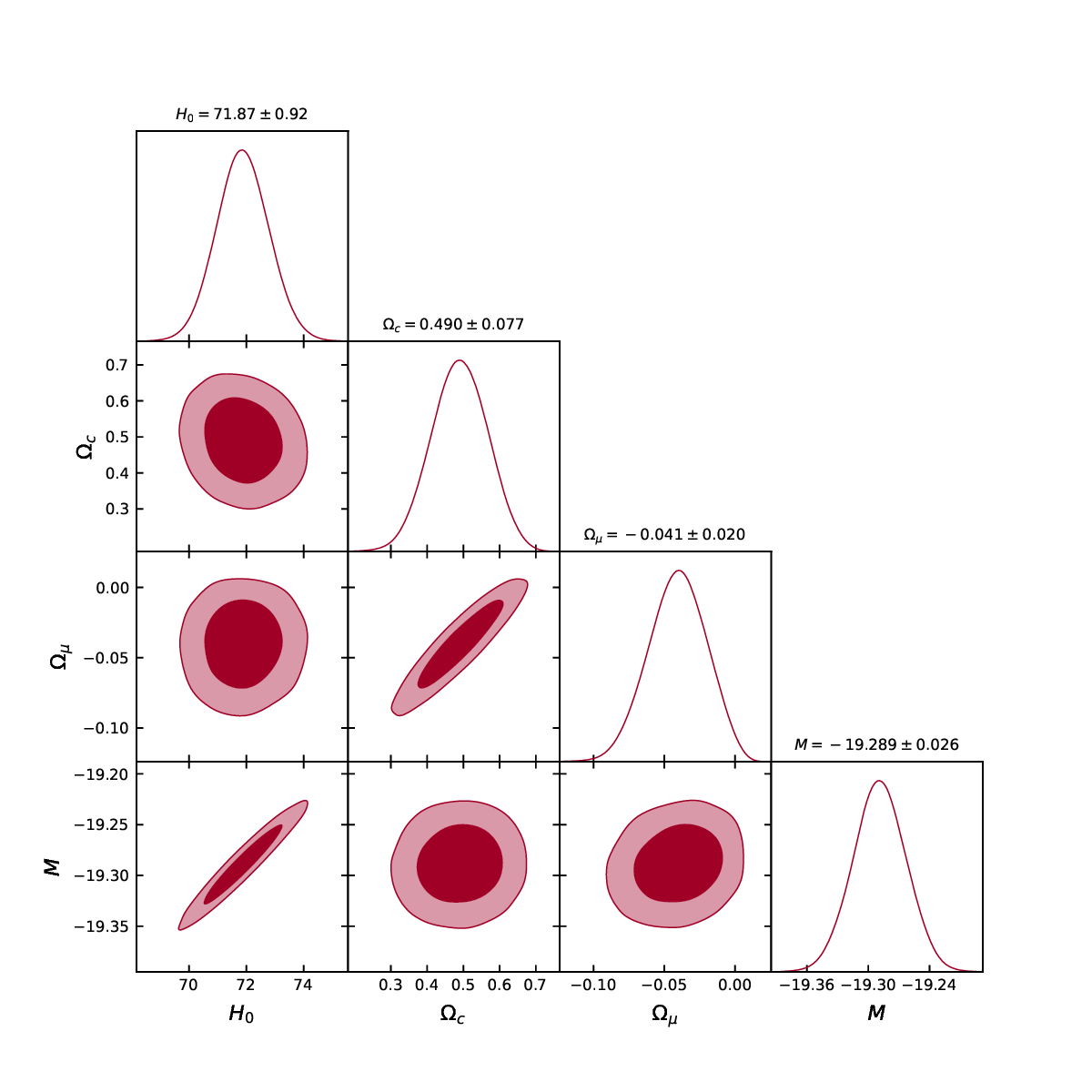}
	\caption{The contour plots of $H_{0}$, $\Omega_{m}$, $\Omega_{\mu}$, $M$ and the contour plots of $H_{0}$, $\Omega_{c}$, $\Omega_{\mu}$, $M$, respectively, for joint analysis of CC$+$Pantheon+SH0ES.}
\end{figure}
\begin{table}[H]
	\centering
	\begin{tabular}{|c|c|c|c|}
		\hline
		Model    & $\rho_{i0}$ (in $gm/cm^{3}$)       & $\mu$ (in $s^2$)         & $\nu$ (in $s^{-2}$) \\
		\hline
		I        & $2.5819\times10^{-38}$            & $2.5548\times10^{33}$       & $3.2765\times10^{-35}$  \\
		\hline
		II       & $1.1110\times10^{-37}$            & $1.3314\times10^{32}$       & $2.7285\times10^{-35}$  \\
		\hline
		III      & $2.0467\times10^{-37}$            & $-2.3389\times10^{35}$       & $2.2333\times10^{-35}$  \\
		\hline
		IV       & $3.1714\times10^{-37}$            & $-6.6307\times10^{36}$       & $1.5258\times10^{-35}$  \\
		\hline
	\end{tabular}
	\caption{The values of other model parameters for CC$+$Pantheon+SH0ES datasets.}\label{T2}
\end{table}
Utilizing the minimum $\chi^{2}$ value (maximizing the likelihood function), we conduct MCMC analysis on the CC$+$Pantheon+SH0ES datasets for Models I, II, III, and IV to estimate the model parameters associated with their respective Hubble functions. The contour plots of the estimated parameters at $1-\sigma$ and $2-\sigma$ confidence levels are illustrated in Figures 1a, 1b, 2a, and 2b. The estimated values of the model parameters are presented in Table \ref{T1}. Utilizing the estimated values of Table \ref{T1}, we assessed the approximated values of other model parameters, which are shown in Table \ref{T2}. By using the priors shown in Table \ref{T1} and running $48$ walkers and $6000$ iterations, we were able to estimate that the constrained values for Models I and II are $H_{0}=73.54\pm0.96, 72.24\pm0.92$ Km/s/Mpc. We have measured the value of the absolute magnitude as $M=-19.317\pm0.027$ for Model I and $M=-19.307\pm0.026$ for Model II. Using the priors as in Table \ref{T1} with $48$ walkers and $10000$ iterations for estimation of model parameters of Models III and IV, we have measured the Hubble constant $H_{0}=72.01\pm0.92, 71.87\pm0.92$ Km/s/Mpc, respectively. We have measured the value absolute magnitude $M=-19.295\pm0.026$ for Model III, while for Model IV $M=-19.289\pm0.026$. For the $\Lambda$CDM model, we used the priors shown in Table \ref{T1} along with $64$ walkers and $12000$ iterations to estimate the model parameters $H_{0}$, $\Omega_{m}$, and $M$. We got $H_{0}=72.02\pm0.91$ Km/s/Mpc, $\Omega_{m}=0.316\pm0.020$, and $M=-19.295\pm0.025$. The absolute magnitude $M$ may be variable and it can be constrained by Pantheon+SH0ES datasets as worked out in \cite{ref90,ref91,ref92,ref93,ref94}. Recently, \cite{ref97} has measured the value of this absolute magnitude $M=-19.294\pm0.026$ using CC+Pantheon+SH0ES datasets.\\

\subsection{Information Criteria}

In this part, we employ the Akaike Information Criterion (AIC) and the Bayesian Information Criterion (BIC) as model selection criteria to statistically compare our derived models fitting with the standard cosmological $\Lambda$CDM model. The AIC criteria can be computed using the following formula, as given in references \cite{ref98,ref99}, under the assumption of Gaussian errors.
\begin{equation}\label{eq56}
	AIC=\chi^{2}_{\text{tot}, \text{min}}+\frac{2kN_{\text{tot}}}{N_{\text{tot}}-k-1},
\end{equation}
The symbol $\chi^{2}_{\text{tot}, \text{min}}$ reflects the minimum value of the $\chi^{2}$ for joint analysis of CC$+$Pantheon+SH0ES datasets being studied. The variable $N_{\text{tot}}$ represents the total number of data points utilized in the analysis, whereas $k$ indicates the number of parameters that have been fitted in a particular model. Minimizing the $\chi^{2}$-value is comparable to maximizing the likelihood function $\mathcal{L}$. The BIC criteria can be computed using the formula specified in references \cite{ref98,ref99,ref100}.
\begin{equation}\label{eq57}
	BIC=\chi^{2}_{\text{tot}, \text{min}}+k\ln{N_{\text{tot}}}
\end{equation}
Our objective is to categorize the models based on their capacity to effectively match the provided data, while considering a range of scenarios that depict the same type of event. When evaluating a group of models, we detect bias in the information criteria (IC) value. To do this, we utilize two specific ICs: AIC and BIC. The expression $\Delta{IC_{model}}=IC_{model}-IC_{min}$ represents the difference between the IC value of the derived model ($IC_{model}$) and the minimal IC value corresponding to the standard model ($IC_{min}$). To evaluate the suitability of each model, we employ Jeffrey's scale \cite{ref101}. More precisely, if the value of $\Delta$IC is less than or equal to $2$, it indicates that the data strongly supports the conventional model. When the disparity value, $\Delta$IC, falls within the range of $2$ to $6$, it signifies a substantial level of disagreement between the two models. Ultimately, when the disparities reach or exceed $10$, it signifies a moderate level of tension between the two models \cite{ref102}.\\
\begin{table}[H]
	\centering
	\[
\begin{array}{|c|c|c|c|c|c|c|c|}
	\hline
	\text{Model} & \chi^2_{\text{tot},\text{min}} & N_{\text{tot}} & k & \text{AIC} & \Delta \text{AIC} & \text{BIC} & \Delta \text{BIC} \\[0.1cm]
	\hline
	\Lambda \text{CDM}& 1547.7061 & 1732 & 3 & 1553.7199 & 0        & 1570.0772 & 0 \\[0.1cm]
	\hline
	\text{Model I}    & 1771.3013 & 1732 & 3 & 1777.3151 & 223.5952 & 1793.6724 & 223.5952 \\[0.1cm]
	\hline
	\text{Model II}   & 1581.2280 & 1732 & 3 & 1587.2418 & 33.5219  & 1603.5991 & 33.5219 \\[0.1cm]
	\hline
	\text{Model III}  & 1547.7011 & 1732 & 4 & 1555.7242 & 2.0043   & 1577.5292 & 7.4520 \\[0.1cm]
	\hline
	\text{Model IV}   & 1544.6908 & 1732 & 4 & 1552.7139 & 1.0060   & 1574.5189 & 4.4417 \\[0.1cm]
	\hline
\end{array}
\]
	\caption{The information criteria}\label{T3}
\end{table}
We found the model parameters by fitting them to the CC$+$Pantheon+SH0ES datasets to get the lowest $\chi^{2}$ value. The minimum $\chi^{2}$ value for each model is shown in Table \ref{T1}. Here, the total number of datasets is denoted by $N_{\text{tot}}=1732$ for the CC$+$Pantheon+ datasets, and the number of fitted parameters with CC$+$Pantheon+ datasets are $k=3, 3, 4, 4$ for Models I, II, III, and IV, respectively. In the case of the standard $\Lambda$CDM model, $k=3$ for CC$+$Pantheon+ datasets. Using these values in the formulas \eqref{eq56} and \eqref{eq57}, we have determined the values of AIC, BIC, $\Delta$AIC, and $\Delta$BIC, which are presented in Table \ref{T3}.

\section{Result discussions}

In this part, we present a comparative study of the analysis of all four derived universe models, I, II, III, and IV. The Hubble function has been determined for each model in terms of $z$ and free parameters. Subsequently, the best-fit values of these free parameters have been estimated using MCMC analysis with joint analysis of $31$ cosmic chronometer (CC) Hubble datasets and $1701$ Pantheon+SH0ES dataset. The minimum $\chi^{2}$ values and values of the fitted parameters along with their corresponding confidence levels of $1-\sigma$ and $2-\sigma$, are presented in Table \ref{T1}. As a result, we have obtained the AIC, $\Delta$AIC, and BIC, $\Delta$BIC values for each model fitting (see Table \ref{T3}). We have calculated the effective equation of the state parameter, denoted as $\omega_{eff}$, and the deceleration parameter for each model to explore the physical characteristics of the models. \\

Based on the information provided in Table \ref{T1}, it is evident that the minimum $\chi^{2}$ value for Model I is very distant, whereas for Model III, it is equivalent to the least $\chi^{2}$ value for the $\Lambda$CDM. We found that the Hubble constants for Models I, II, III, and IV are $H_{0}=73.54\pm0.96,~72.24\pm0.92,~72.01\pm0.92,~71.87\pm0.92$ Km/s/Mpc. The estimated values of some other model parameters are shown in Table \ref{T4}. Furthermore, from Table \ref{T3}, it becomes evident that the differences in AIC and BIC values for Model I, II exceed $10$, indicating a mild tension between Model I, II and $\Lambda$CDM. The $\Delta$AIC value for Model III falls in the range $(2, 6)$ and $\Delta$BIC value for Model III falls in the range of $(6, 10)$, which means that it shows less tension with the $\Lambda$CDM in comparison of Models I, II. For the model IV, the $\Delta$IC value for the falls in the range of $(1, 2)$, which means that this model is closed to $\Lambda$CDM while the $\Delta$BIC value of Model IV falls in $(2, 6)$ which shows some tension between Model IV and $\Lambda$CDM. Thus, the Models III, and IV are substantially closed to $\Lambda$CDM standard model statistically while there is a mild tension between Model I, II and $\Lambda$CDM.\\
\begin{figure}[H]
	\centering
	\includegraphics[width=10cm,height=8cm,angle=0]{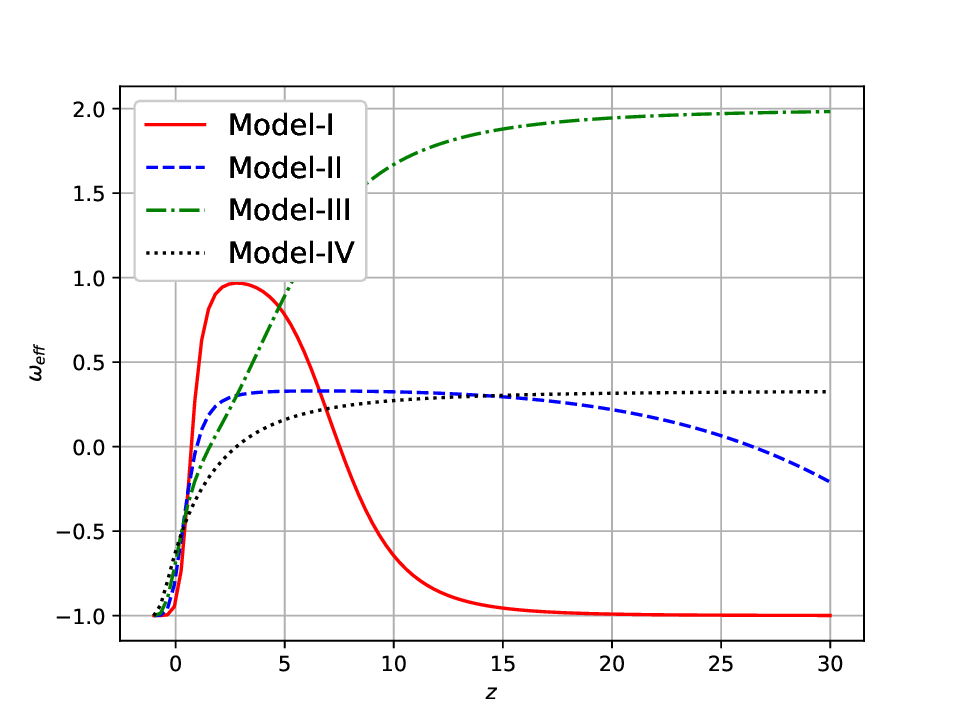}
	\caption{The variations of effective EoS parameter $\omega_{eff}$ versus $z$.}
\end{figure}

Figure 3 displays the variations of the effective equation of state parameter $\omega_{eff}$ for Models I, II, III, and IV across different redshift values over $-1\le z \le 30$. The mathematical formulas for these parameters are provided in equations \eqref{eq28}, \eqref{eq34}, \eqref{eq40}, and \eqref{eq47}, respectively. Figure 3 shows that the effective equation of state (EoS) parameter for Models I and II starts out negative in the early stages of the universe. It then goes up to its highest positive value, then goes down and slowly approaches $-1$ as the universe gets older. On the other hand, the equation of state parameter $\omega_{eff}$ for Models III and IV initially has a positive value and approaches $-1$ in the later stages of the universe. The calculated present value of the effective equation of state (EoS) parameter is determined for models I, II, III, and IV, and these are shown in Table \ref{T4}. One can observe that the present values of effective EoS parameter falls in the range $-0.9238\le\omega_{eff0}\le-0.6186$ which are compatible with recent estimations in different investigations. From Table \ref{T4}, we can see that Model I, II have higher dark energy as $0.8301\le\Omega_{\nu0}\le0.9618$ while in Model III and IV $0.5510\le\Omega_{\nu0}\le0.6838$ which may in good agreement with standard $\Lambda$CDM value $\Omega_{\Lambda0}=0.70$. Figures 3 and 4 demonstrate that models I, II, III, and IV exhibit accelerated expansion within the specified range of the effective equation of state parameter, $-1\le\omega_{eff}<-\frac{1}{3}$, where model I and II show two phases of accelerating scenarios early and late-time universe. This demonstrates the viability of our generated models. The models I, II, and III, behaves just like dark energy late-time accelerating cosmological models for the barotropic fluid source $0\le\omega_{i}\le1$.\\

\begin{table}[H]
	\centering
	\begin{tabular}{|c|c|c|c|c|}
		\hline
		
		Model    & $\omega_{eff0}$       & $q_{0}$         & $z_{t}$       & $\Omega_{\nu0}$\\
		\hline
		I        & $-0.9238$            & $-0.8857$       & $8.5570$      & $0.9618$\\
		&                      &                 & $0.5259$      & \\
		\hline
		II       & $-0.7734$            & $-0.6602$       & $31.6232$     & $0.8301$\\
		&                      &                 & $0.4867$      & \\
		\hline
		III      & $-0.6814$            & $-0.5221$       & $0.5780$      & $0.6838$\\
		\hline
		IV       & $-0.6186$            & $-0.4279$       & $0.8390$      & $0.5510$\\
		\hline
	\end{tabular}
	\caption{The value of parameters.}\label{T4}
\end{table}

\begin{figure}[H]
	\centering
	\includegraphics[width=10cm,height=8cm,angle=0]{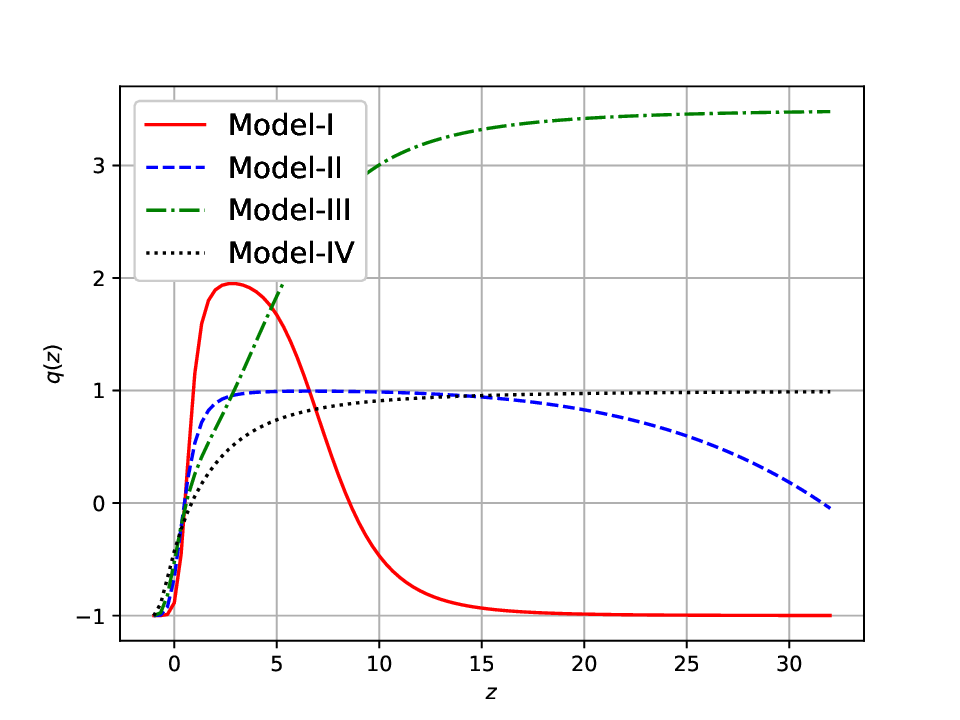}
	\caption{The variations of deceleration parameter $q(z)$ versus $z$.}
\end{figure}

The estimated present values of the energy density parameters $\Omega_{s}$, $\Omega_{r}$, $\Omega_{m}$, $\Omega_{c}$, and $\Omega_{\mu}$ are shown in Table \ref{T1}, while the present values of the dark energy density parameter $\Omega_{\nu}$ are shown in Tables \ref{T4}. Based on these measurements, we can see that universe models I and II, which have stiff and radiation fluid sources, have a lot more dark energy from the $f(R, L_{m}, T)$ function in the form of $\Omega_{\nu}$ than universe models III and IV. The second part of the Lagrangian function $f(R,L_{m},T)$ makes a lot more of a difference in the last two models, III and IV, compared to the first two models, I and II. The non-zero values of $\Omega_{\mu}$ and $\Omega_{\nu}$ represent the significant role of modified terms in Einstein's Lagrangian. You can see the estimated values of the model parameters $\rho_{i0} (i=s, r, m, c)$, $\mu$, and $\nu$ in Table \ref{T2}.\\

The mathematical formulas for the deceleration parameter are given in Eqs. \eqref{eq29}, \eqref{eq35}, \eqref{eq41}, and \eqref{eq48} for Models I, II, III, and IV, respectively. Figure 4 show how the deceleration parameter $q(z)$ changes shape for Models I, II, III, and IV, using the estimated values of the model parameters along observational datasets. It is evident that the plot of $q(z)$ for Models I and II exhibits a negative value at the beginning and at the end within the redshift range of $-1\le z \le 32$ and within this range once $q(z)$ reaches its maximum positive value. This observation indicates that Models I and II show two phases: transitioning from accelerating to decelerating (early stage) and then decelerating to accelerating (late-time scenarios) in each model. On the other hand, $q(z)$ for Models III and IV shows transition from decelerating to accelerating phase. The estimated present values of $q(z)$ for each model are shown in Tables \ref{T4}, which are as follows: $q_{0}=-0.8857, -0.6602, -0.5221, -0.4279$. The transition redshift $z_{t}$ is measured to be $z_{t}=0.5259, 0.4867, 0.5780, 0.8390$ for Models I, II, III, and IV, respectively, for late-time transition, while Models I and II show a transition in its early stages at $z_{t}=8.557, 31.6232$, respectively. Therefore, it can be concluded that all of the generated models are transit phase universes that experience late-time acceleration during the expansion of the universe, while Models I and II show two phases of acceleration and deceleration. The value of the transition redshift was recently estimated to be $z_{t}=0.8596_{-0.2722}^{+0.2886}$ for SNIa datasets, and the value of $z_{t}=0.6320_{-0.1403}^{+0.1605}$ is obtained for Hubble datasets in $f(R)$ gravity models \cite{ref103}. The research \cite{ref104} presents a transition redshift of $z_{t}=0.643_{-0.030}^{+0.034}$ in the framework of $f(T)$ gravity. The transition value $z_t$ was measured by \cite{ref105} using a model-independent technique as $z_{t}=0.646_{-0.158}^{+0.020}$. The transition value reported in another study by \cite{ref106} is $z_{t}=0.702_{-0.044}^{+0.094}$. In addition, the transition redshift value was measured by \cite{ref107,ref108} to be $z_{t}=0.684_{-0.092}^{+0.136}$. Therefore, our estimated values of $z_{t}$ for Models I, II, III, and IV align with the aforementioned values.\\

\subsection{Causality}

To analyze the stability of derived cosmological models, we investigate the square sound speed $c_{s}^{2}$ defined as \cite{ref109}
\begin{equation}\label{eq58}
	c_{s}^{2}=\frac{dp_{e}}{d\rho_{e}}
\end{equation}
where $p_{e}=p_{eff}$ and $\rho_{e}=\rho_{eff}$. For a perfect fluid source the square speed of sound can not exceed the square velocity of light i.e., $c_{s}^{2}\le c^{2}$ \cite{ref109}. Hence, for a stable cosmological model, the square speed of sound should be less than unity (for $c=1$), i.e., $0\le c_{s}^{2}\le1$. Thus, we can derive the speed of sound for effective pressure and energy density of each model using equation \eqref{eq58}, respectively, as follows:
\begin{equation}\label{eq59}
	c_{s}^{2}=\frac{\Omega_{s}-2\Omega_{\mu}(1+z)^{6}}{\Omega_{s}+2\Omega_{\mu}(1+z)^{6}}
\end{equation}
\begin{equation}\label{eq60}
	c_{s}^{2}=\frac{1}{3}-\frac{8}{3}\frac{\Omega_{\mu}}{\Omega_{r}}(1+z)^{4}
\end{equation}
\begin{equation}\label{eq61}
	c_{s}^{2}=\frac{4\Omega_{\mu}(1+z)^{3}}{\Omega_{m}+2\Omega_{\mu}(1+z)^{3}}
\end{equation}
\begin{equation}\label{eq62}
	c_{s}^{2}=\frac{-\Omega_{c}-2\Omega_{\mu}(1+z)^{2}}{3\Omega_{c}-6\Omega_{\mu}(1+z)^{2}}
\end{equation}
The equations \eqref{eq59}, \eqref{eq60}, \eqref{eq61}, and \eqref{eq62} represent the expressions of square sound speed for the models I, II, III, and IV, respectively, in terms of energy density parameters and redshift $z$. Figure 5 depicts the evolution of square sound speed $c_{s}^{2}$ versus redshift $z$. One can see that the models for barotropic fluid have square sound speed values within the given range $0\le c_{s}^{2}\le1$, whereas model IV violates this condition in late-time stages. Therefore, the models I, II, and III are stable, but model IV is unstable at late-time.
\begin{figure}[H]
	\centering
	\includegraphics[width=10cm,height=8cm,angle=0]{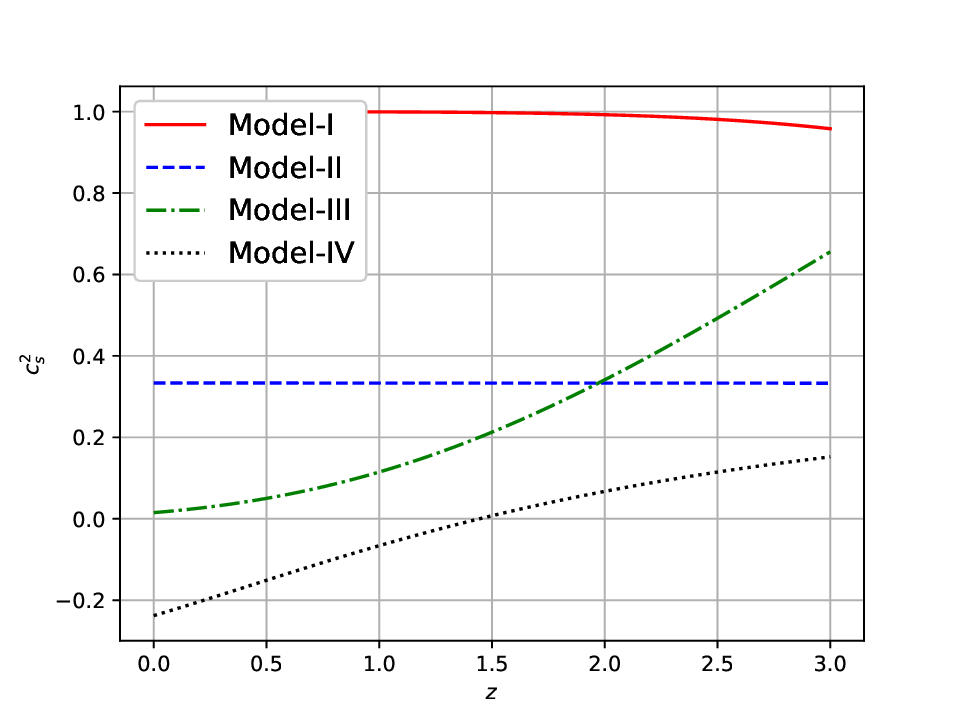}
	\caption{The plot of square sound speed $c_{s}^{2}$ versus $z$.}
\end{figure}

\subsection{$f(R,L_{m},T)$ function}

Within this part, we analyze the Lagrangian function $f(R,L_{m},T)$. In order to accomplish this, we can derive the expression for the function $f(R,L_{m},T)$ for each model separately, as follows:
\begin{equation}\label{eq63}
	f_{I}(R,L_{m},T)=-16\pi\rho_{s0}(1+z)^{6}-10\mu\rho_{s0}^{2}(1+z)^{12}+\nu,
\end{equation}

\begin{equation}\label{eq64}
	f_{II}(R,L_{m},T)=\nu,
\end{equation}

\begin{equation}\label{eq65}
	f_{III}(R,L_{m},T)=8\pi\rho_{m0}(1+z)^{3}+2\mu\rho_{m0}^{2}(1+z)^{6}+\nu,
\end{equation}

\begin{equation}\label{eq66}
	f_{IV}(R,L_{m},T)=16\pi\rho_{c0}(1+z)^{2}+2\mu\rho_{c0}^{2}(1+z)^{4}+\nu.
\end{equation}
The partial derivatives of the function $f(R,L_{m},T)$ with respect to $L_{m}$ and $T$ are calculated and presented in Table \ref{T5} for each model.
\begin{table}[H]
	\centering
	\begin{tabular}{|c|c|c|c|}
		\hline
		
		Models      & $f_{L_{m}}=\mu T$              & $f_{T}=\mu L_{m}$    & Value of $\mu$\\
		\hline
		I           & $2\mu\rho$                     & $-\mu\rho$           & $\mu>0$\\
		\hline
		II          & $0$                            & $-\mu\rho$           & $\mu>0$ or $\mu<0$\\ 
		\hline
		III         & $-\mu\rho$                     & $-\mu\rho$           & $\mu<0$\\
		\hline
		IV          & $-2\mu\rho$                    & $-\mu\rho$           & $\mu<0$\\
		\hline
	\end{tabular}
	\caption{The derivatives of function $f(R,L_{m},T)$.}\label{T5}
\end{table}

By examining equations \eqref{eq63}-\eqref{eq66}, it becomes evident that the function $f(R,L_{m},T)$ approaches $\nu$ as $z$ approaches $-1$, with $\nu$ being greater than zero. Conversely, $f(R,L_{m},T)$ approaches negative infinity as $z$ approaches infinity, regardless of the model. Furthermore, according to the information provided in Table \ref{T5}, it can be observed that $f_{L_{m}}\ge0$ for all models while $f_{T}>0$ for models II, III, and IV but $f_{T}<0$ for model I. This demonstrates the feasibility of our derived universe models.

\section{Conclusions}

In the generalized matter-geometry coupling theory, we investigated the physical characteristics and causality of some new cosmological models for a flat, homogeneous, and isotropic spacetime filled with stiff, radiation, dust, and curvature fluid sources. We have obtained a particular cosmological model corresponding to each source fluid, called Models I, II, III, and IV, respectively. We have made observational constraints on each model using joint analysis of the $31$ CC Hubble dataset and $1701$ Pantheon+SH0ES datasets to estimate the current values of model parameters. Using these statistical results, we have analyzed the information criteria, effective EoS parameter, causality of the models, and viability of this generalized gravity theory. We have measured the values of the Hubble constant in the range $71.87\le H_{0} \le 73.54$, which are compatible with recent observed values. We have found that all the derived models are transit phase-accelerating in late-time universes, while models I and II also depict the accelerating phase in their early stages. We have obtained the current values of the deceleration parameter $q(z)$ in the range $-0.8857\le q_{0}\le -0.4279$ with the transition redshift $0.4867\le z_{t}\le0.8390$ for all Models I, II, III, and IV, in their late-time stages, while Models I and II have a transition point from accelerating to decelerating in the transition redshift range $8.557\le z_{t}\le31.6232$ (for detail see Table \ref{T4}). We have measured the value of absolute magnitude in the range $-19.317\le M \le-19.289$ which is in good agreement with recent estimations in \cite{ref90,ref91,ref92,ref93,ref94,ref97}.\\

We observed that the effective equation of state (EoS) parameter for Models I and II initially had a negative value during the early stages of the universe, then increased to a maximum positive value, and then decreased to $-1$ in the later stages of the universe. On the other hand, the equation of state parameter $\omega_{eff}$ for Models III and IV initially has a positive value and approaches $-1$ in the later stages of the universe. We found the present value of the effective EoS parameter to be $-0.9238\le\omega_{eff}\le-0.6186$. We have concluded that all models exhibit an accelerated expansion phase within the specified range of the effective equation of state parameter, $-1\le\omega_{eff}<-\frac{1}{3}$, whereas models I and II show two phases of accelerating scenarios: early and late-time universes. The models I, II, and III behave just like dark energy late-time accelerating cosmological models for the barotropic fluid source $0\le\omega_{i}\le1$. We have measured the present values of $\Omega_{\mu}$, $\Omega_{\nu}$, $\mu$, $\nu$, and $\rho_{i0} (i=s, r, m, c)$, which are mentioned in Tables \ref{T1}, \ref{T2}, \ref{T4}. The non-zero values of $\Omega_{\mu}$ and $\Omega_{\nu}$ represent the significant role of modified terms in Einstein's Lagrangian.\\

We have found that statistical analysis of joint datasets represents a mild tension between Model I and $\Lambda$CDM in their analysis of AIC and BIC (see Table \ref{T3}). Furthermore, in Models III, and IV, the values of $\Delta$AIC and $\Delta$BIC fall below $10$, suggesting a minor deviation from the $\Lambda$CDM model while the values of $\Delta$AIC and $\Delta$BIC for Models I and II fall upper $10$, suggesting a mild deviation from the $\Lambda$CDM model. We found the square sound speed values for Models I, II, and III within the range $0\le c_{s}^{2}\le1$, which illustrates that they are in stable stages, while Model IV violates this condition of stability. For each model, the Lagrangian function $f(R,L_{m},T)$ and its derivatives $f_{L_{m}}$ and $f_{T}$ have been computed. The positive derivatives $f_{L_{m}}\ge0$ and $f_{T}>0$ indicate that our derived models are feasible. In this way, the cosmological models in the generalized matter-geometry coupling gravity theory can explain the physical features of a universe that can be seen. In general, this theory doesn't meet the energy conditions, and more research is needed to see how the formation or destruction of matter takes place in this $f(R,L_{m},T)$ gravity theory.

\section*{Acknowledgments}

The authors are thankful to renowned and honorable reviewers and editors for their valuable suggestions to improve the quality of this paper. The author Dinesh Chandra Maurya is thankful to IUCAA Center for Astronomy Research and Development (ICARD), CCASS, GLA University, Mathura, India for providing facilities and support where part of this work is carried out.

\section{Data Availability Statement}
No data associated in the manuscript.

\section{Declarations}
\subsection*{Funding and/or Conflicts of interests/Competing interests}
The authors of this article have no conflict of interests. The authors have no competing interests to declare that are relevant to the content of this article. The authors did not receive support from any organization for the submitted work.



\begin{thebibliography}{}
	\bibitem {ref1}
A. G. Riess, A. V. Filippenko, P. Challis, \textit{et al.},
Observational evidence from supernovae for an accelerating universe and a cosmological constant,
\textit{Astron. J.} \textbf{116} (1998) 1009-1038,
[arXiv:astro-ph/9805201]
\bibitem {ref2}
S. Perlmutter, G. Aldering, G. Goldhaber, \textit{et al.},
Measurements of Omega and Lambda from 42 High-Redshift Supernovae,
\textit{Astrophys. J.} \textbf{517} (1999) 565-586,
[arXiv:astro-ph/9812133]
\bibitem {ref3}
R. A. Knop, G. Aldering, R. Amanullah, \textit{et al.},
New Constraints on $\Omega_{M}$, $\Omega_{\Lambda}$, and $\omega$ from an Independent Set of 11 High-Redshift Supernovae Observed with the Hubble Space Telescope, 
\textit{Astrophys. J.} \textbf{598} (2003) 102,
[arXiv:astro-ph/0309368]
\bibitem {ref4}
R. Amanullah, C. Lidman, D. Rubin, \textit{et al.},
Spectra and Hubble Space Telescope light curves of six type Ia supernovae at $0.511< z< 1.12$ and the Union2 compilation
\textit{Astrophys. J.} \textbf{716} (2010) 712-738,
[arXiv:1004.1711 [astro-ph.CO]]
\bibitem {ref5}
D. H. Weinberg, M. J. Mortonson, D. J. Eisenstein, \textit{et al.},
Observational probes of cosmic acceleration,
\textit{Phys. Rep.} \textbf{530} (2013) 87-255,
[arXiv:1201.2434 [astro-ph.CO]]
\bibitem {ref6}
A. Einstein, The Principle of General Relativity, \textit{Annalen Physik Leipzig} \textbf{55} (1918) 241-245.
\bibitem {ref7}
P. Salucci, N. Turini, and C. Di Paolo,
Paradigms and scenarios for the dark matter phenomenon,
\textit{Universe} \textbf{6} (2020) 118,
[arXiv:2008.04052 [astro-ph.CO]]
\bibitem {ref8}
S. Alam, M. Ata, S. Bailey, \textit{et al.} (BOSS Collaboration),
The clustering of galaxies in the completed SDSS-III Baryon Oscillation Spectroscopic Survey: cosmological analysis of the DR12 galaxy sample,
\textit{Mon. Not. R. Astron. Soc.} \textbf{470} (2017) 2617-2652,
[arXiv:1607.03155 [astro-ph.CO]]
\bibitem {ref9}
T. M. C. Abbott, F. B. Abdalla, A. Alarcon, \textit{et al.} (DES Collaboration),
Dark Energy Survey year 1 results: Cosmological constraints from galaxy clustering and weak lensing,
\textit{Phys. Rev. D} \textbf{98} 043526 (2018),
[arXiv:1708.01530 [astro-ph.CO]]
\bibitem {ref10}
M. Tanabashi, K. Hagiwara, K. Hikasa, \textit{et al.} (Particle Data Group),
Review of Particle Physics: particle data groups,
\textit{Phys. Rev. D} \textbf{98}, 030001 (2018),
\bibitem {ref11}
N. Aghanim, Y. Akrami, M. Ashdown, \textit{et al.} (Planck Collaboration),
Planck 2018 results. VI. Cosmological parameters,
\textit{Astron. Astrophys.} \textbf{641}, A6 (2020),
[arXiv:1807.06209 [astro-ph.CO]]
\bibitem {ref12}
H. A. Buchdahl,
Non-linear Lagrangians and cosmological theory,
\textit{Mon. Not. R. Astron. Soc.} \textbf{150}, 1-8 (1970),
\bibitem {ref13}
R. Kerner,
Cosmology without singularity and nonlinear gravitational Lagrangians,
\textit{Gen. Relativ. Gravit.} \textbf{14} 453-469 (1982),
\bibitem {ref14}
J. P. Duruisseau, R. Kerner, P. Eysseric,
Non-Einsteinian gravitational Lagrangians assuring cosmological solutions without collapse,
\textit{Gen. Relativ. Gravit.} \textbf{15} 797-807 (1983),
\bibitem {ref15}
J. D. Barrow, A. C. Ottewill,
The stability of general relativistic cosmological theory,
\textit{J. Phys. A Math. Gen.} \textbf{16}, 2757 (1983),
\bibitem {ref16}
H. Kleinert, H.-J. Schmidt,
Cosmology with Curvature-Saturated Gravitational Lagrangian R,
\textit{Gen. Relativ. Gravit.} \textbf{34}, 1295-1318 (2002),
[arXiv:gr-qc/0006074]
\bibitem {ref17}
S.M. Carroll, V. Duvvuri, M. Trodden, M.S. Turner,
Is cosmic speed-up due to new gravitational physics?,
\textit{Phys. Rev. D} \textbf{70}, 043528 (2004).
[arXiv:astro-ph/0306438]
\bibitem {ref18}
W. Hu, I. Sawicki,
Models of $f(R)$ cosmic acceleration that evade solar system tests,
\textit{Phys. Rev. D} \textbf{76}, 064004 (2007),
[arXiv:0705.1158 [astro-ph]]
\bibitem {ref19}
S. A. Appleby, R. A. Battye,
Do consistent $F(R)$ models mimic general relativity plus $\Lambda$?,
\textit{Phys. Lett. B} \textbf{654}, 7-12 (2007),
[arXiv:0705.3199 [astro-ph]]
\bibitem {ref20}
A. A. Starobinsky,
Disappearing cosmological constant in $f(R)$ gravity,
\textit{JETP Lett.} \textbf{86}, 157-163 (2007),
[arXiv:0706.2041 [astro-ph]]
\bibitem {ref21}
V. Faraoni,
de Sitter space and the equivalence between $f(R)$ and scalar-tensor gravity,
\textit{Phys. Rev. D} \textbf{75}, 067302 (2007),
[arXiv:gr-qc/0703044]
\bibitem {ref22}
C. G. B\"{o}hmer, L. Hollenstein, F. S. N. Lobo,
Stability of the Einstein static universe in $f(R)$ gravity,
\textit{Phys. Rev. D} \textbf{76}, 084005 (2007),
[arXiv:0706.1663 [gr-qc]]
\bibitem {ref23}
C. G. B\"{o}hmer, T. Harko, F. S. N. Lobo,
Dark matter as a geometric effect in $f(R)$ gravity,
\textit{Astropart. Phys.} \textbf{29}, 386-392 (2008),
[arXiv:0709.0046 [gr-qc]]
\bibitem {ref24}
S. A. Appleby, R. A. Battye, A. A. Starobinsky,
Curing singularities in cosmological evolution of $F(R)$ gravity,
\textit{JCAP} \textbf{1006}, 005 (2010),
[arXiv:0909.1737 [astro-ph.CO]]
\bibitem {ref25}	
S. Chakraborty, K. MacDevette, P. Dunsby,
A Model independent approach to the study of $f(R)$ cosmologies with expansion histories close to $\Lambda$CDM,
\textit{Phys. Rev. D} \textbf{103}, 124040 (2021),
[arXiv:2103.02274 [gr-qc]]
\bibitem {ref26}
V. K. Oikonomou,
Unifying inflation with early and late dark energy epochs in axion $F(R)$ gravity,
\textit{Phys. Rev. D} \textbf{103}, 044036 (2021),
[arXiv:2012.00586 [astro-ph.CO]]
\bibitem{ref27}
S.~Nojiri and S.~D.~Odintsov,
Modified gravity with negative and positive powers of the curvature: Unification of the inflation and of the cosmic acceleration,
\textit{Phys. Rev. D} \textbf{68} (2003), 123512
[arXiv:hep-th/0307288 [hep-th]].
\bibitem{ref28}
S.~Nojiri and S.~D.~Odintsov,
Modified f(R) gravity consistent with realistic cosmology: From matter dominated epoch to dark energy universe,
\textit{Phys. Rev. D} \textbf{74} (2006), 086005
[arXiv:hep-th/0608008 [hep-th]].
\bibitem{ref29}
G.~Cognola, E.~Elizalde, S.~Nojiri, S.~D.~Odintsov, L.~Sebastiani and S.~Zerbini,
A Class of viable modified f(R) gravities describing inflation and the onset of accelerated expansion,
\textit{Phys. Rev. D} \textbf{77} (2008), 046009
[arXiv:0712.4017 [hep-th]].
\bibitem{ref30}
S.~Nojiri and S.~D.~Odintsov,
Unified cosmic history in modified gravity: from F(R) theory to Lorentz non-invariant models,
\textit{Phys. Rept.} \textbf{505} (2011), 59-144
[arXiv:1011.0544 [gr-qc]].
\bibitem{ref31}
S.~Nojiri and S.~D.~Odintsov,
Unified cosmic history in modified gravity: from F(R) theory to Lorentz non-invariant models,
\textit{Phys. Rept.} \textbf{505} (2011), 59-144
[arXiv:1011.0544 [gr-qc]].
\bibitem {ref32}
A. Dixit, D. C. Maurya, A. Pradhan,
Transit cosmological models coupled with zero-mass scalar field with high redshift in higher derivative theory,
\textit{New Astronomy} \textbf{87}, 101587 (2021),
[arXiv:2005.11139 [physics.gen-ph]]
\bibitem {ref33}
A. Pradhan, De Avik, Tee How Loo, D. C. Maurya,
A flat FLRW model with dynamical $\Lambda$ as function of matter and geometry,
\textit{New Astronomy} \textbf{89}, 101637 (2021),
[arXiv:2010.09444 [physics.gen-ph]]
\bibitem {ref34}
D. C. Maurya, R. Zia, A. Pradhan,
Anisotropic string cosmological model in Brans–Dicke theory of gravitation with time-dependent deceleration parameter,
\textit{J. Exp. Theor. Phys.} \textbf{123}, 617-622 (2016),
\bibitem {ref35}
D. C. Maurya, R. Zia, A. Pradhan,
Dark energy models in LRS Bianchi type-II space-time in the new perspective of time-dependent deceleration parameter,
\textit{Int. J. Geom. Meth. Mod. Phys.} \textbf{14}(05) 1750077 (2017),
\bibitem {ref36}
D. C. Maurya,
Anisotropic dark energy transit cosmological models with time-dependent $\omega(t)$ and redshift-dependent $\omega(z)$ EoS parameter,
\textit{Int. J. Geom. Meth. Mod. Phys.} \textbf{15}(02) 1850019 (2018),
\bibitem {ref37}
R. Zia, D. C. Maurya,
Brans–Dicke scalar field cosmological model in Lyra’s geometry with time-dependent deceleration parameter,
\textit{Int. J. Geom. Meth. Mod. Phys.} \textbf{15}(11) 1850186 (2018),
\bibitem {ref38}
D. C. Maurya, R. Zia,
Brans-Dicke scalar field cosmological model in Lyra’s geometry,
\textit{Physical Review D} \textbf{100}(2) 023503 (2019),
[arXiv:1907.07135 [gr-qc]]
\bibitem {ref39}
R. Zia, U. K. Sharma, D. C. Maurya,
Transit two-fluid models in anisotropic Bianchi type-III space-time,
\textit{New Astronomy} \textbf{72}, 83-91 (2019),
\bibitem {ref40}
D. C. Maurya, R. Zia,
Reply to ``Comment on `Brans-Dicke scalar field cosmological model in Lyra's geometry'",
\textit{Phys. Rev. D} \textbf{102}(10) 108302 (2020),
\bibitem {ref41}
O. Bertolami, C. G. Boehmer, T. Harko, F. S. N. Lobo,
Extra force in $f(R)$ modified theories of gravity,
\textit{Phys. Rev. D} \textbf{75}, 104016 (2007),
[arXiv:0704.1733 [gr-qc]]
\bibitem {ref42}
T. P. Sotiriou, V. Faraoni,
$f(R)$ theories of gravity,
\textit{Rev. Mod. Phys.} \textbf{82}, 451 (2010),
[arXiv:0805.1726 [gr-qc]]
\bibitem {ref43}
A. De Felice, S. Tsujikawa,
$f(R)$ theories,
\textit{Living Rev. Relativ.} \textbf{13}, 3 (2010),
[arXiv:1002.4928 [gr-qc]]
\bibitem {ref44}
T. Harko,
Modified gravity with arbitrary coupling between matter and geometry,
\textit{Phys. Lett. B} \textbf{669}, 376 (2008),
[arXiv:0810.0742 [gr-qc]]
\bibitem {ref45}
T. Harko, F. S. N. Lobo,
$f(R,L_{m})$ gravity,
\textit{Eur. Phys. J. C} \textbf{70}, 373 (2010),
[arXiv:1008.4193 [gr-qc]]
\bibitem {ref46}
T. Harko,
The matter Lagrangian and the energy-momentum tensor in modified gravity with nonminimal coupling between matter and geometry,
\textit{Phys. Rev. D} \textbf{81}, 044021 (2010),
[arXiv:1001.5349 [gr-qc]]
\bibitem {ref47}
T. Harko, F. S. N. Lobo,
Geodesic deviation, Raychaudhuri equation, and tidal forces in modified gravity with an arbitrary curvature-matter coupling,
\textit{Phys. Rev. D} \textbf{86}, 124034 (2012),
[arXiv:1210.8044 [gr-qc]]
\bibitem {ref48}
J. Wang, K. Liao,
Energy conditions in $f(R, L_{m})$ gravity,
\textit{Class. Quantum Gravity} \textbf{29}, 215016 (2012),
[arXiv:1212.4656 [physics.gen-ph]]
\bibitem {ref49}
O. Minazzoli, T. Harko,
New derivation of the Lagrangian of a perfect fluid with a barotropic equation of state,
\textit{Phys. Rev. D} \textbf{86}, 087502 (2012),
[arXiv:1209.2754 [gr-qc]]
\bibitem {ref50}
T. Harko, F. S. N. Lobo, O. Minazzoli,
Extended $f(R,L_{m})$ gravity with generalized scalar field and kinetic term dependences,
\textit{Phys. Rev. D} \textbf{87}, 047501 (2013),
[arXiv:1210.4218 [gr-qc]]
\bibitem {ref51}
D. W. Tian, I. Booth,
Lessons from $(R,R_{c}^{2},R_{m}^{2},L_{m})$ gravity: Smooth Gauss-Bonnet limit, energy-momentum conservation, and nonminimal coupling,
\textit{Phys. Rev. D} \textbf{90}, 024059 (2014),
[arXiv:1404.7823 [gr-qc]]
\bibitem {ref52}
T. Harko, F. S. N. Lobo, J. P. Mimoso, D. Pav\'{o}n,
Gravitational induced particle production through a nonminimal curvature–matter coupling,
\textit{Eur. Phys. J. C} \textbf{75}, 386 (2015),
[arXiv:1508.02511 [gr-qc]]
\bibitem {ref53}
R. P. L. Azevedo, P. P. Avelino,
Big-bang nucleosynthesis and cosmic microwave background constraints on nonminimally coupled theories of gravity,
\textit{Phys. Rev. D} \textbf{98}, 064045 (2018),
[arXiv:1807.00798 [gr-qc]]
\bibitem {ref54}
S. Bahamonde,
Generalised nonminimally gravity-matter coupled theory,
\textit{Eur. Phys. J. C} \textbf{78}, 326 (2018),
[arXiv:1709.05319 [gr-qc]]
\bibitem {ref55}
R. March, O. Bertolami, M. Muccino, \textit{et al.},
Constraining a nonminimally coupled curvature-matter gravity model with ocean experiments,
\textit{Phys. Rev. D} \textbf{100}, 042002 (2019),
[arXiv:1904.12789 [gr-qc]]
\bibitem {ref56}
O. Bertolami, C. Gomes,
Nonminimally coupled Boltzmann equation: foundations,
\textit{Phys. Rev. D} \textbf{102}, 084051 (2020),
[arXiv:2002.08184 [gr-qc]]
\bibitem{ref57}
G.~Allemandi, A.~Borowiec, M.~Francaviglia and S.~D.~Odintsov,
Dark energy dominance and cosmic acceleration in first order formalism,
\textit{Phys. Rev. D} \textbf{72} (2005), 063505
[arXiv:gr-qc/0504057 [gr-qc]].
\bibitem {ref58}
A. Pradhan, D. C. Maurya, G. K. Goswami, A. Beesham,
Modeling Transit Dark Energy in $F(R, L_{m})$-gravity,
\textit{Int. J. Geom. Meth. Mod. Phys.}, \textbf{20}(06) 2350105 (2023),
[arXiv:2209.14269 [gr-qc]]
\bibitem {ref59}
D. C. Maurya,
Accelerating scenarios of massive universe in $f(R, L_{m})$-gravity,
\textit{New Astronomy} \textbf{100}, 101974 (2023),
\bibitem {ref60}
D. C. Maurya,
Exact Cosmological Models in Modified $F(R,L_{m})$-Gravity with Observational Constraints,
\textit{Grav. \& Cosmo.} \textbf{29}(3) 315-325 (2023),
\bibitem {ref61}
D. C. Maurya,
Bianchi-I dark energy cosmological model in $f(R,L_{m})$-Gravity,
\textit{Int. J. Geom. Meth. Mod. Phys.} \textbf{21}(04) 2450072-194 (2024),
\bibitem {ref62}
D. C. Maurya,
Constrained $\Lambda$CDM Dark Energy Models in Higher Derivative $F(R,L_{m})$-Gravity Theory,
\textit{Phys. Dark Universe} \textbf{42} (2023) 101373,
\bibitem {ref63}
T. Harko, F. S. N. Lobo, S. Nojiri, S. D. Odintsov,
$f(R,T)$ gravity,
\textit{Phys. Rev. D} \textbf{84}, 024020 (2011),
[arXiv:1104.2669 [gr-qc]]
\bibitem {ref64}
T. Harko,
Thermodynamic interpretation of the generalized gravity models with geometry-matter coupling,
\textit{Phys. Rev. D} \textbf{90}, 044067 (2014),
[arXiv:1408.3465 [gr-qc]]
\bibitem {ref65}
E. H. Baffou, M. J. S. Houndjo, M. E. Rodrigues, \textit{et al.},
Cosmological evolution in $f(R,T)$ theory with collisional matter,
\textit{Phys. Rev. D} \textbf{92}, 084043 (2015),
[arXiv:1504.05496 [gr-qc]]
\bibitem {ref66}
T. B. Gonçalves, J. L. Rosa, F. S. N. Lobo,
Cosmology in scalar-tensor $f(R,T)$ gravity,
\textit{Phys. Rev. D} \textbf{105}, 064019 (2022),
[arXiv:2112.02541 [gr-qc]]
\bibitem {ref67}
H. Shabani, A.H. Ziaie,
Stability of the Einstein static universe in $f(R, T)$ gravity,
\textit{Eur. Phys. J. C} \textbf{77}, 31 (2017),
[arXiv:1606.07959 [gr-qc]]
\bibitem {ref68}
H. Velten, T. R. P. Caram\^{e}s,
Cosmological inviability of $f(R,T)$ gravity,
\textit{Phys. Rev. D} \textbf{95}, 123536 (2017),
[arXiv:1702.07710 [gr-qc]]
\bibitem {ref69}
P. H. R. S. Moraes, R. A. C. Correa, G. Ribeiro,
Evading the non-continuity equation in the $f(R, T)$ cosmology,
\textit{Eur. Phys. J. C} \textbf{78}, 192 (2018).
\bibitem {ref70}
S. K. Maurya, A. Errehymy, Ksh. Newton Singh, \textit{et al.},
Gravitational decoupling minimal geometric deformation model in modified $f(R, T)$ gravity theory,
\textit{Phys. Dark Universe} \textbf{30}, 100640 (2020).
\bibitem {ref71}
T. Harko, P. H. R. S. Moraes, 
Comment on ``Reexamining $f(R,T)$ gravity",
\textit{Phys. Rev. D} \textbf{101}, 108501 (2020).
\bibitem {ref72}
R. Zia, D. C. Maurya, A. Pradhan,
Transit dark energy string cosmological models with perfect fluid in $F(R,T)$-gravity,
\textit{Int. J. Geom. Meth. Mod. Phys.} \textbf{15}(10) 1850168 (2018).
\bibitem {ref73}
D. C. Maurya,
Modified $F(R,T)$ cosmology with observational constraints in Lyra’s geometry,
\textit{Int. J. Geom. Meth. Mod. Phys.} \textbf{17}(01) 2050001 (2020).
\bibitem {ref74}
D. C. Maurya, A. Pradhan, A. Dixit,
Domain walls and quark matter in Bianchi type-V universe with observational constraints in $F(R,T)$ gravity,
\textit{Int. J. Geom. Meth. Mod. Phys.} \textbf{17}(01) 2050014 (2020).
\bibitem {ref75}
D. C. Maurya,
Transit cosmological model with specific Hubble parameter in $F(R,T)$ gravity,
\textit{New Astronomy} \textbf{77}, 101355 (2020).
\bibitem {ref76}
D. C. Maurya, J. singh, L. K. Gaur,
Dark Energy Nature in Logarithmic $f(R,T)$ Cosmology,
\textit{Int. J. Geom. Meth. Mod. Phys.} \textbf{20}(11) 2350192 (2023).
\bibitem {ref77}
G. P. Singh, N. Hulke, A. Singh,
Cosmological study of particle creation in higher derivative theory,
\textit{Indian J. Phys.} \textbf{94} 127-141 (2020).
\bibitem {ref78}
N. Hulke, G. P. Singh, B. K. Bishi, A. Singh,
Variable Chaplygin gas cosmologies in $f(R, T)$ gravity with particle creation,
\textit{New Astronomy} \textbf{77} 101357 (2020).
\bibitem {ref79}
Z Haghani and T. Harko,
Generalizing the coupling between geometry and matter: $f(R, L_{m}, T)$ gravity,
\textit{Eur. Phys. J. C} (2021) \textbf{81}:615.
\bibitem {ref80}
S. Arora, P.H.R.S. Moraes, and P.K. Sahoo, 
Energy conditions in the $f(R, L, T)$ theory of gravity,
\textit{Eur. Phys. J. Plus} \textbf{139} (2024) 542.
\bibitem {ref81}
A. Pradhan, M. Zeyauddin, A. Dixit, and S. Krishnannair,
Dust-fluid accelerating flat cosmological models in $f(R,T,L_{m})$-gravity with observational constraints,
\textit{Int. J. Geom. Meth. Mod. Phys.}, (2024),
https://doi.org/10.1142/S0219887825500604.
\bibitem {ref82}
J.S. Gonçalves, A.F. Santos,
On causality and its violation in $f(R,L_{m},T)$ gravity,
\textit{Nuclear Physics B} \textbf{1010} (2025) 116751.
\bibitem {ref83}
D. C. Maurya,
Late-time accelerating cosmological models in $f(R,L_{m},T)$-gravity with observational constraints,
\textit{Phys. Dark Universe} \textbf{46} (2024) 101722.
\bibitem {ref84}
D. C. Maurya,
Constrained transit cosmological models in $f(R,L_{m},T)$-gravity,
\textit{Int. J. Geom. Meth. Mod. Phys.}, (2024),
https://doi.org/10.1142/S0219887825500288.
\bibitem{ref85}
S. Capozziello, O. Luongo, R. Pincak, \textit{et al.},
Cosmic acceleration in non-flat $f(T)$ cosmology,
\textit{Gen. Relativ. Gravit.} \textbf{50}, 53 (2018),
[arXiv:1804.03649v1 [gr-qc]]. 
\bibitem {ref86}
L. D. Landau, E. M. Lifshitz,The Classical Theory of Fields (Butterworth-Heinemann, Oxford, 1998).
\bibitem{ref87}
E. J. Copeland, M. Sami and S. Tsujikawa,
Dynamics of dark energy,
\textit{Int. J. Mod. Phys. D} \textbf{15}, 1753 (2006),
[arXiv:hep-th/0603057].
\bibitem {ref88}
J. Simon, L. Verde, R. Jimenez,
Constraints on the redshift dependence of the dark energy potential,
\textit{Phys. Rev. D} \textbf{71}, 123001 (2005)
\bibitem {ref89}
G. S. Sharov, V. O. Vasiliev,
How predictions of cosmological models depend on Hubble parameter data sets,
\textit{Math. Model. Geom.} \textbf{6}, 1-20 (2018)
\bibitem{ref90}
D. Brout, \textit{et al.}, The Pantheon+ Analysis: Cosmological Constraints, \textit{ApJ} \textbf{938} 110 (2022).
\bibitem{ref91}
D. Scolnic, \textit{et al.}, The Pantheon+ Analysis: The Full Data Set and Light-curve Release, \textit{ApJ} \textbf{938} 113 (2022).
\bibitem{ref92}
L. Perivolaropoulos, F. Skara, On the homogeneity of SnIa absolute magnitude in the Pantheon+ sample, \textit{MNRAS} \textbf{520} 5110-5125 (2023), https://doi.org/10.1093/mnras/stad451.
\bibitem{ref93}
Adam G. Riess \textit{et al.}, A Comprehensive Measurement of the Local Value of the Hubble Constant with $1\, km s^{-1} Mpc^{-1}$ Uncertainty from the Hubble Space Telescope and the SH0ES Team, \textit{ApJL} \textbf{934} L7 (2022). [arXiv:2112.04510 [astro-ph.CO]].
\bibitem{ref94}
http://PantheonPlusSH)ES.github.io.
\bibitem{ref95}
D.W. Hogg and D.F. Mackey,
Data analysis recipes: Using Markov Chain Monte Carlo,
\textit{The Astrophysical Journal Supplement Series} \textbf{236} (2018) 18.
[arXiv:1710.06068 [astro-ph.IM]].
\bibitem{ref96}
D. M. Scolnic et al., The Complete Light-curve Sample of Spectroscopically Confirmed SNe Ia from Pan-STARRS1 and Cosmological Constraints from the Combined Pantheon Sample, \textit{ApJ} \textbf{859} (2018) 101.
\bibitem{ref97}
A. Singh, S. Mandal, R. Chaubey, R. Raushan,
Observational constraints on the expansion scalar and shear relation in the Locally rotationally symmetric Bianchi I model,
\textit{Phys. Dark Univ.} \textbf{47} (2025) 101798,
https://doi.org/10.1016/j.dark.2024.101798.
\bibitem {ref98}
K. Anderson, Model selection and multimodel inference: a practical information-theoretic approach, 2nd edn. Springer, New York (2002).
\bibitem {ref99}
K. P. Burnham, D. R. Anderson,
Multimodel Inference: Understanding AIC and BIC in Model Selection,
\textit{Sociological Methods and Research} \textbf{33}, 261 (2004).
\bibitem {ref100}
A. R. Liddle,
Information criteria for astrophysical model selection,
\textit{Mon. Not. Roy. Astron. Soc.} \textbf{377}, (2007) L74,
[arXiv:astro-ph/0701113].
\bibitem {ref101}
R. E. Kass and A. E. Raftery,
Bayes Factors,
\textit{J. Am. Statist. Assoc.} \textbf{90} 773 (1995).
\bibitem {ref102}
F. K. Anagnostopoulos, S. Basilakos, E.N. Saridakis,
Observational constraints on Myrzakulov gravity,
\textit{Phys. Rev. D} \textbf{103}, 104013 (2021),
[arXiv:2012.06524 [gr-qc]].
\bibitem{ref103}
S. Capozziello, O. Farooq, O. Luongo, \textit{et al.},
Cosmographic bounds on the cosmological deceleration-acceleration transition redshift in $f(R)$ gravity,
\textit{Phys. Rev. D} \textbf{90}, 044016 (2014),
[arXiv:1403.1421v1 [gr-qc]].
\bibitem{ref104}
S. Capozziello, O. Luongo, E. N. Saridakis,
Transition redshift in $f(T)$ cosmology and observational constraints,
\textit{Phys. Rev. D} \textbf{91} 124037 (2015),
[arXiv:1503.02832v2 [gr-qc]].
\bibitem{ref105}
S. Capozziello, P. K. S. Dunsby, O. Luongo,
Model independent reconstruction of cosmological accelerated-decelerated phase,
\textit{MNRAS} \textbf{509} (2022) 5399-5415,
[arXiv:2106.15579v2 [astro-ph.CO]].
\bibitem{ref106}
M. Muccino, O. Luongo, D. Jain,
Constraints on the transition redshift from the calibrated Gamma-ray Burst Ep-Eiso correlation,
\textit{MNRAS} \textbf{523} (2023) 4938-4948,
[arXiv:2208.13700v3 [astro-ph.CO]].
\bibitem{ref107}
A. C. Alfano, S. Capozziello, O. Luongo, \textit{et al.},
Cosmological transition epoch from gamma-ray burst correlations, (2024) 
[arXiv:2402.18967v1 [astro-ph.CO]].
\bibitem{ref108}
A. C. Alfano, C. Cafaro, S. Capozziello, \textit{et al.},
Dark energy-matter equivalence by the evolution of cosmic equation of state,
\textit{Phys. Dark Univ.} \textbf{42} 101298 (2023),
[arXiv:2306.08396v2 [astro-ph.CO]].
\bibitem{ref109}
G. Ellis, R. Maartens, M. MacCallum,
Causality and the speed of sound,
\textit{Gen. Rel. Grav.} \textbf{39} 1651-1660 (2007),
[arXiv:gr-qc/0703121]

\end{thebibliography}
\end{document}